\documentclass[aps,prl,showpacs,twocolumn,superscriptaddress]{revtex4-1}

\usepackage{graphicx,amsmath,amssymb}
\usepackage[usenames]{color}
\usepackage{float}
\usepackage[T1]{fontenc}

\usepackage[colorlinks=true, citecolor=blue, urlcolor=blue, linkcolor=blue ]{hyperref}
\hypersetup{breaklinks=true}
\begin{document}

\bibliographystyle{apsrev4-1}

\title{Theory of Kondo hybridization wave in Kondo lattice}
\author{Yin Zhong}
\email{zhongy@lzu.edu.cn}
\affiliation{Key Laboratory of Quantum Theory and Applications of MoE $\&$ School of Physical Science and Technology, Lanzhou University, Lanzhou 730000, People Republic of China}
\affiliation{Lanzhou Center for Theoretical Physics, Key Laboratory of Theoretical Physics of Gansu Province, Lanzhou University, Lanzhou 730000, People Republic of China}
\begin{abstract}
Recent scanning tunneling microscopy experiments have discovered emergent spatially modulated Kondo hybridization wave (KHW) order in the heavy fermion superconductor UTe$_2$ and the artificial Kondo lattice system 1T/1H-TaS$_2$, challenging the conventional paradigm of spatially uniform Kondo hybridization in heavy fermion physics. Here, we develop a microscopic theory for KHW order based on the canonical square-lattice Kondo lattice model within the large-$N$ fermionic mean-field approximation. We systematically identify stable modulated KHW phases and establish their ground-state phase diagram. The prominent $\boldsymbol{Q}=(0,\pi)$ KHW phase yields uniaxial stripe modulation of the Kondo hybridization gap, which faithfully reproduces the spatial modulation pattern observed in UTe$_2$. Moreover, its inherent unit-cell-doubling modulation precisely accounts for the spectroscopic features measured in 1T/1H-TaS$_2$. We further predict a characteristic in-plane conductivity anisotropy that serves as a definitive transport fingerprint to discriminate KHW states with distinct ordering wavevectors. Our work provides a microscopic foundation for the newly observed KHW order and establishes a unified theoretical framework for understanding emergent modulated hybridization phenomena in heavy fermion materials.
\end{abstract}

\maketitle
\emph{Introduction}.--
A single magnetic impurity embedded in a metallic Fermi sea gives rise to the paradigmatic low-temperature Kondo effect\cite{Hewson1993}. When magnetic moments form a periodic lattice, as realized in heavy-fermion materials\cite{Doniach1977}, local Kondo screening evolves into coherent hybridization between itinerant conduction electrons and localized $f$-electrons\cite{Lohneysen2007}. Conventionally, Kondo hybridization is assumed to be spatially uniform. This uniform hybridization scenario constitutes the foundation of the standard heavy-fermion hybridized band picture and provides the central framework for describing Kondo-breakdown quantum criticality beyond the Landau-Ginzburg-Wilson paradigm\cite{Lohneysen2007,Coleman2015,Coleman2001,Si2001,Senthil2004,Vojta2010}.

Above conventional scenario has been fundamentally challenged by recent scanning tunneling microscopy (STM) experiments on the heavy fermion superconductor UTe$_{2}$ and the artificial Kondo lattice system 1T/1H-TaS$_{2}$\cite{Yu2026,Cao2026}. These measurements unambiguously reveal spatially modulated Kondo hybridization, establishing an emergent Kondo hybridization wave (KHW) order absent from standard theoretical descriptions. In particular, the STM-resolved local density of states exhibits a uniaxial unit-cell-doubling modulation of the hybridization gap, a feature that cannot be captured by existing phenomenological two-band models or conventional Ginzburg-Landau analyses\cite{Dubi2011,Su2011}. We emphasize that such intrinsic oscillatory hybridization is distinct from extrinsic Friedel-like oscillations induced by local Kondo holes in low-dimensional Kondo-Heisenberg systems\cite{Xie2017}. Despite these striking experimental observations, a comprehensive microscopic theory for KHW order remains lacking. The absence of lattice model framework hinders the microscopic understanding of its spectral, thermodynamic, and transport properties, and further limits its potential relevance to long-standing open problems such as the hidden-order phase in URu$_{2}$Si$_{2}$\cite{Dubi2011,Su2011,Schmidt2010,Mydosh2011}.

In this work, we resolve this theoretical gap by developing a fully microscopic description of KHW states within the canonical Kondo lattice model. Departing from the conventional spatially uniform hybridization ansatz, we allow both the Kondo hybridization order parameter and the Lagrange multiplier to host momentum-resolved, spatially oscillatory components. Within the large-$N$ mean-field framework, we stabilize a rich set of KHW ordered phases and establish their ground-state phase diagram. Notably, the characteristic $\boldsymbol{Q}=(0,\pi)$ KHW phase naturally yields uniaxial unit-cell doubling of the hybridization gap, quantitatively consistent with STM observations on 1T/1H-TaS$_{2}$. Meanwhile, its stripe-like hybridization modulation extracted from Fano line-shape of calculated STM spectrum reproduces the key spatial pattern observed in UTe$_{2}$\cite{Morr2017,Yang2009,Maltseva2009,Morr2010,Wolfle2010}. Furthermore, we predict pronounced in-plane conductivity anisotropy as a characteristic transport signature to differentiate distinct KHW wavevector states, providing an experimentally accessible route for future verification in heavy-fermion materials.

\begin{figure}
\includegraphics[width=0.95\linewidth]{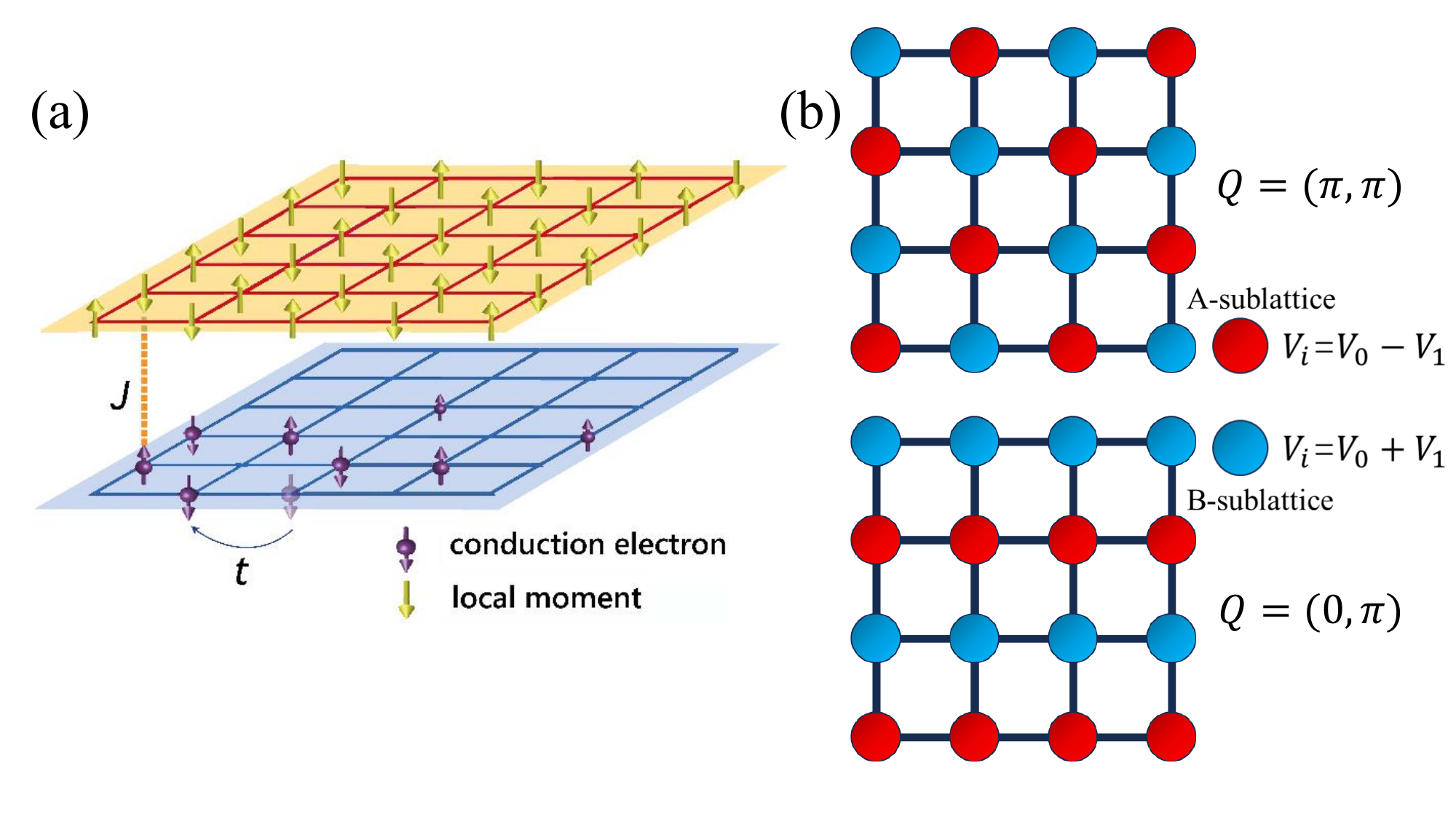}
\caption{\label{fig:1} (a) The Kondo lattice model defined on square lattice with nearest-neighbor-hopping $t$ and Kondo coupling $J$. (b) Schematic spatial distribution of Kondo hybridization order parameter $V_{i}=V_{0}+V_{1}e^{iQ\cdot R_{i}}$ with $Q=(\pi,\pi)$ and $(0,\pi)$.}
\end{figure}
\emph{Model}.--
The Kondo lattice model defined on square lattice is\cite{Coleman2015,Tsunetsugu1997}
\begin{equation}
H=\sum_{k\sigma}\varepsilon_{k}c_{k\sigma}^{\dag}c_{k\sigma}+J\sum_{i}\vec{S}_{i}^{c}\cdot\vec{S}_{i}^{f}.
\label{eq1}
\end{equation}
Here, $c_{k\sigma}^{\dag}$ is the creation operator for a conduction electron carrying momentum $k$ and spin index $\sigma=\uparrow,\downarrow$.
The conduction-electron dispersion on the square lattice is given by $\varepsilon_{k}=-2t(\cos k_{x}+\cos k_{y})-4t'\cos k_{x}\cos k_{y}-\mu$ with
$t$, $t'$ and $\mu$ denoting, respectively, the nearest-neighbor-hopping, next-nearest-neighbor-hopping and chemical potential. Other lattice geometry can be implemented analogously. Through this work, we fix $t=1$ to serve as our energy unit and set $t'=0$. We restrict ourselves to the on-site Kondo coupling $J$, even though nonlocal extensions of this interaction have been reported in literature\cite{Ghaemi2007,Weber2008,Coleman1998}. (see Fig.~\ref{fig:1}(a))

Using the standard Abrikosov fermion (parton) representation for $S=\frac{1}{2}$ local moment of $f$-electron  $\vec{S}_{i}^{f}=\frac{1}{2}\sum_{\sigma\sigma'}f^{\dag}_{i\sigma}\vec{\sigma}_{\sigma\sigma'}f_{i\sigma'}$ subject to the single-occupancy constraint $\sum_{\sigma}f_{i\sigma}^{\dag}f_{i\sigma}=1$ imposed on each lattice site,
we have $\vec{S}_{i}^{c}\cdot\vec{S}_{i}^{f}=-\frac{1}{2}\sum_{\sigma\sigma'}c_{i\sigma}^{\dag}f_{i\sigma}f_{i\sigma'}^{\dag}c_{i\sigma'}+\frac{1}{4}\sum_{\sigma} c_{i\sigma}^{\dag}c_{i\sigma}$\cite{Coleman2015,Lacroix1979,Zhang2011,Irkhin2023}. The latter term $\frac{1}{4}\sum_{\sigma} c_{i\sigma}^{\dag}c_{i\sigma}$ can be absorbed into the redefinition of chemical potential $\mu$.
After the mean-field decoupling $c_{i\sigma}^{\dag}f_{i\sigma}f_{i\sigma'}^{\dag}c_{i\sigma'}
\simeq
\langle c_{i\sigma}^{\dag}f_{i\sigma}\rangle f_{i\sigma'}^{\dag}c_{i\sigma'}
+c_{i\sigma}^{\dag}f_{i\sigma}\langle f_{i\sigma'}^{\dag}c_{i\sigma'}\rangle
-\langle c_{i\sigma}^{\dag}f_{i\sigma}\rangle\langle f_{i\sigma'}^{\dag}c_{i\sigma'}\rangle$
and introducing the Kondo hybridization/screening order parameter as $V_{i}^{\sigma}=\langle c_{i\sigma}^{\dag}f_{i\sigma}\rangle$, the Kondo interaction term is rewritten as
$H_{K}=J\sum_{i}\vec{S}_{i}^{c}\cdot\vec{S}_{i}^{f}
=-\frac{J}{2}\sum_{i,\sigma\sigma'}[V_{i}^{\sigma}f_{i\sigma'}^{\dag}c_{i\sigma'}
+(V_{i}^{\sigma'})^{\ast}c_{i\sigma}^{\dag}f_{i\sigma}
-V_{i}^{\sigma}(V_{i}^{\sigma'})^{\ast}]$.

Since spatially uniform $V_{i}$ is incapable of stabilizing to KHW order, we parameterize $V_{i}$ as the superposition of a homogeneous background and a spatially modulated contribution, i.e. $V_{i}=V_{0}+V_{1}e^{i\boldsymbol{Q}\cdot\boldsymbol{R}_{i}}$ with $\boldsymbol{Q}$ labeling the characteristic ordering wavevector. ($V_{0},V_{1}$ are set to be real to simplify the formalism)
The wavevector $\boldsymbol{Q}=(\pi,\pi)$ yields a checkerboard-type configuration for $V_{i}$, which doubles the original unit cell into two-sublattice structure (A and B sublattice), reminiscent of the N\'{e}el antiferromagnetic order. (see Fig.~\ref{fig:1}(b)) By contrast, $\boldsymbol{Q}=(0,\pi)$ produces stripe-like order along $x$-direction, breaking both translation along $y$-direction and $C_{4}^{z}$-rotation symmetry of the square lattice.
The resulting Kondo interaction term reads as
\begin{eqnarray}
H_{K}&=&-JV_{0}\sum_{k\sigma}\left[f_{k\sigma}^{\dag}c_{k\sigma}+h.c.\right]-JV_{1}\sum_{k\sigma}\left[f_{k+Q\sigma}^{\dag}c_{k\sigma}+h.c.\right]\nonumber\\
&&+2JN_{s}(V_{0}^{2}+V_{1}^{2}+2V_{0}V_{1}\delta_{Q=0}),\label{eq2}
\end{eqnarray}
where $N_{s}$ is the number of unit-cell for square lattice. Similar to $V_{i}$, the Lagrange multiplier $\lambda_{i}$ is also nonuniform and has the same $\boldsymbol{Q}$-dependence as $V_{i}$, i.e. $\lambda_{i}=\lambda\frac{e^{i\boldsymbol{Q}\cdot \boldsymbol{R_{i}}}+e^{-i\boldsymbol{Q}\cdot \boldsymbol{R_{i}}}}{2}$, so the Lagrange multiplier term is written as
\begin{eqnarray}
H_{\lambda}&=&\sum_{i}\lambda_{i}\left(\sum_{\sigma}f_{i\sigma}^{\dag}f_{i\sigma}-1\right)\nonumber\\
&=&\frac{\lambda}{2}\sum_{k\sigma}(f_{k\sigma}^{\dag}f_{k+Q\sigma}+h.c.)-N_{s}\lambda\delta_{Q=0}.\label{eq3}
\end{eqnarray}
We emphasize that the \emph{$\boldsymbol{Q}$-dependent Kondo hybridization $V_{i}$ and Lagrange multiplier $\lambda_{i}$} are the main assumption of the present work and they inevitably lead to KHW as shown below. Inserting Eqs.~\ref{eq2} and \ref{eq3} into Eq.~\ref{eq1} leads to the following mean-field Hamiltonian
\begin{eqnarray}
H=\sum_{k\sigma}\Psi_{k\sigma}^{\dag}H_{k}\Psi_{k\sigma}+N_{s}\left[2J(V_{0}^{2}+V_{1}^{2}+2V_{0}V_{1}\delta_{Q=0})-\lambda\delta_{Q=0}\right]\nonumber
\end{eqnarray}
with $\Psi_{k\sigma}^{\dag}=(c^{\dag}_{k\sigma}, f^{\dag}_{k+Q\sigma}, c^{\dag}_{k+Q\sigma}, f^{\dag}_{k\sigma})$ and Bloch Hamiltonian
\begin{equation}
H_{k}=\frac{1}{2}\left(
                          \begin{array}{cccc}
                            \varepsilon_{k} & -JV_{1} & 0 & -JV_{0} \\
                            -JV_{1} & 0 & -JV_{0} & \lambda \\
                            0 & -JV_{0} & \varepsilon_{k+Q} & -JV_{1} \\
                            -JV_{0} & \lambda & -JV_{1} & 0 \\
                          \end{array}
                        \right).\label{eq4}
\end{equation}
The mean-field parameters $V_{0},V_{1},\lambda$ are obtained by numerically solving self-consistent equations,
$V_{0}=\frac{1}{N_{s}}\sum_{k}\langle c_{k\sigma}^{\dag}f_{k\sigma}\rangle$,
$V_{1}=\frac{1}{N_{s}}\sum_{k}\langle c_{k\sigma}^{\dag}f_{k+Q\sigma}\rangle$,
$1=\frac{1}{N_{s}}\sum_{k\sigma}\langle f_{k\sigma}^{\dag}f_{k\sigma}\rangle$ and
$n_{c}=\frac{1}{N_{s}}\sum_{k\sigma}\langle c_{k\sigma}^{\dag}c_{k\sigma}\rangle$, wherein the last one fixes the chemical potential for a specific conduction electron density $n_{c}$. All thermodynamic quantities are derived from the free energy density
$F=-\frac{T}{N_{s}}\sum_{k\sigma,n=1,2,3,4}\ln(1+e^{-\beta E_{kn}})+2J(V_{0}^{2}+V_{1}^{2}+2V_{0}V_{1}\delta_{Q=0})-\lambda\delta_{Q=0}$
with $E_{kn}$ being four eigenenergy of Bloch Hamiltonian $H_{k}$ (Eq.~\ref{eq4}).

\emph{Global phase diagram}.--
\begin{figure}
\includegraphics[width=0.95\linewidth]{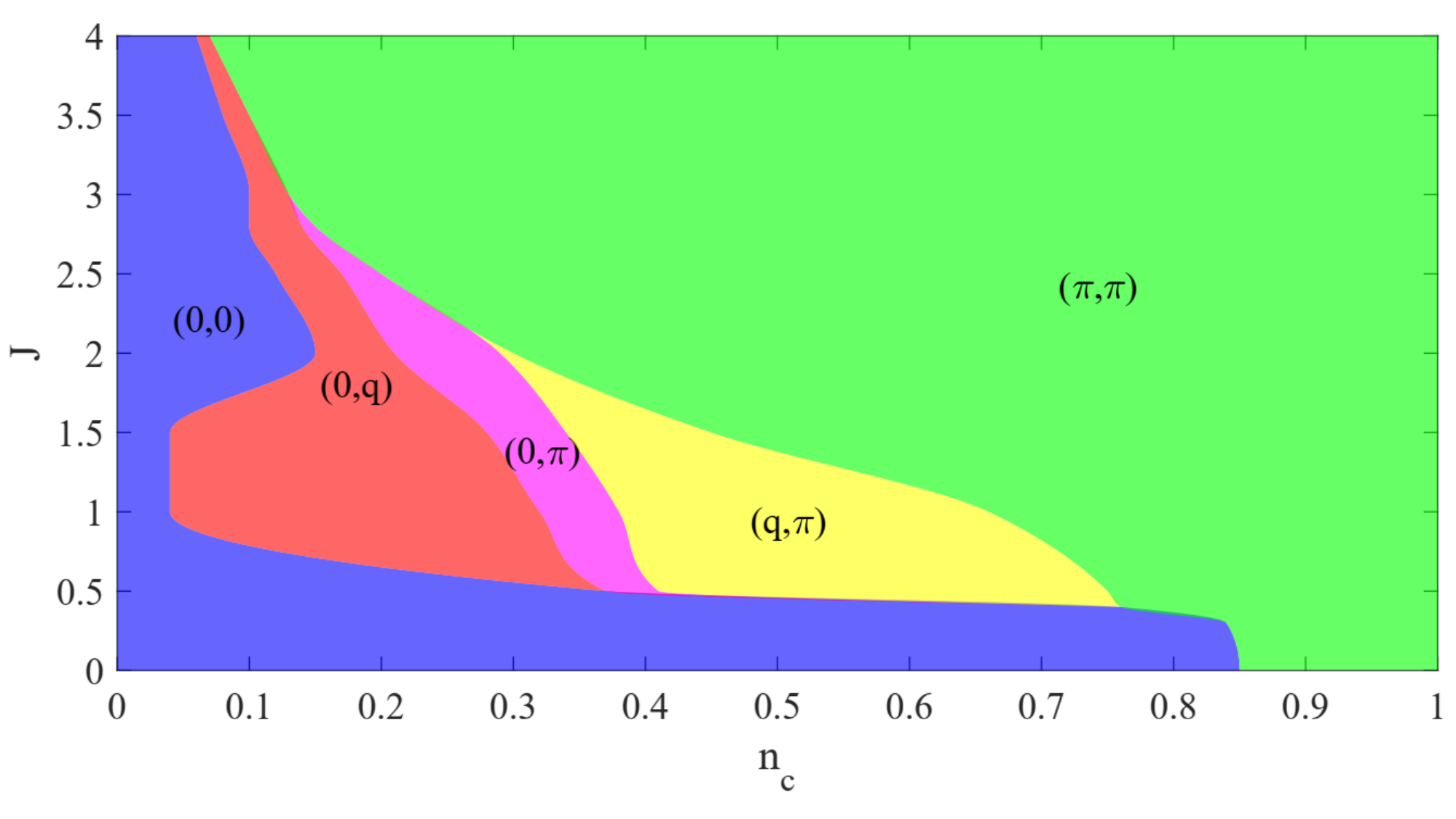}
\caption{\label{fig:2} The $J-n_{c}$ ground-state phase diagram on square lattice with $\boldsymbol{Q}=(Q_{x},Q_{y})$ and $q\in(0,\pi)$.}
\end{figure}
Solving self-consistent equations and comparing free energy
density $F$ for all possible characteristic wavevector $\boldsymbol{Q}=(Q_{x},Q_{y})$ spanning the entire Brillouin zone,
we map out the ground-state phase diagram presented in Fig.~\ref{fig:2}. (see Supplementary Materials (SM) for details and note its similarity to magnetic phase diagram of Kondo lattice\cite{Costa2017}) The $\boldsymbol{Q}=(\pi,\pi)$ KHW phase dominates a wide parameter regime when $J$ is sufficiently large or the conduction electron filling $n_{c}$ lies near half-filling. At low carrier density $n_{c}$, the system favors standard homogeneous Kondo-hybridized state with $\boldsymbol{Q}=(0,0)$. Commensurate KHW solutions with $\boldsymbol{Q}=(0,\pi)$ also exist and occupy finite regions in parameter space, whereas incommensurate KHW phases prevail over the rest of the phase diagram. We concentrate hereafter on $\boldsymbol{Q}=(0,\pi)$ KHW phase, with others discussed in SM.

\begin{figure}
\includegraphics[width=0.95\linewidth]{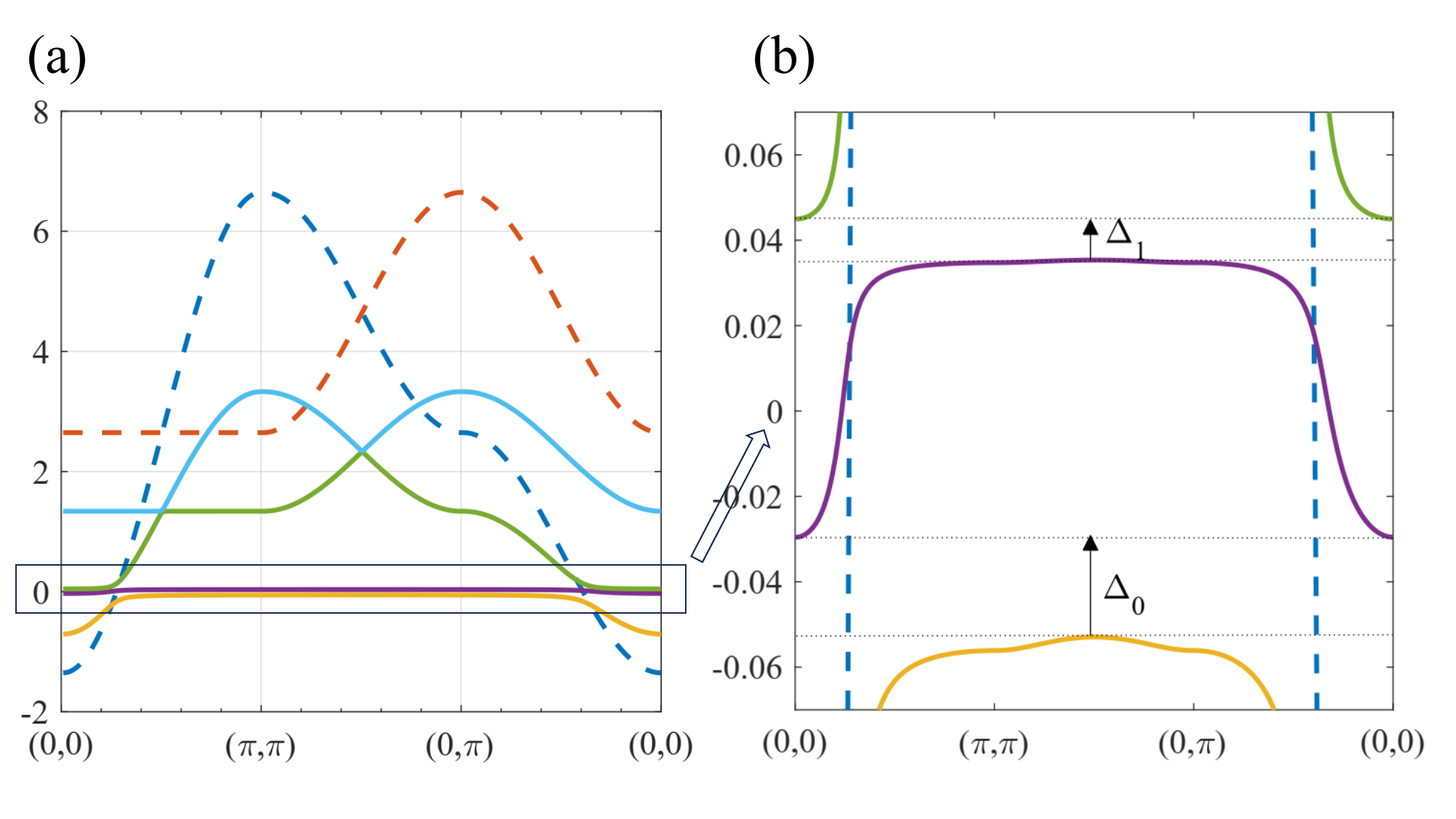}
\caption{\label{fig:3} (a) The quasi-particle bands (solid line and dashed line corresponds to bare bands $\varepsilon_{k},\varepsilon_{k+Q}$) along high-symmetry path $(0,0)-(\pi,\pi)-(0,\pi)-(0,0)$ and (b) their hybridization gap $\Delta_{0},\Delta_{1}$ for KHW phase with $\boldsymbol{Q}=(0,\pi)$.}
\end{figure}

\begin{figure}
\includegraphics[width=0.95\linewidth]{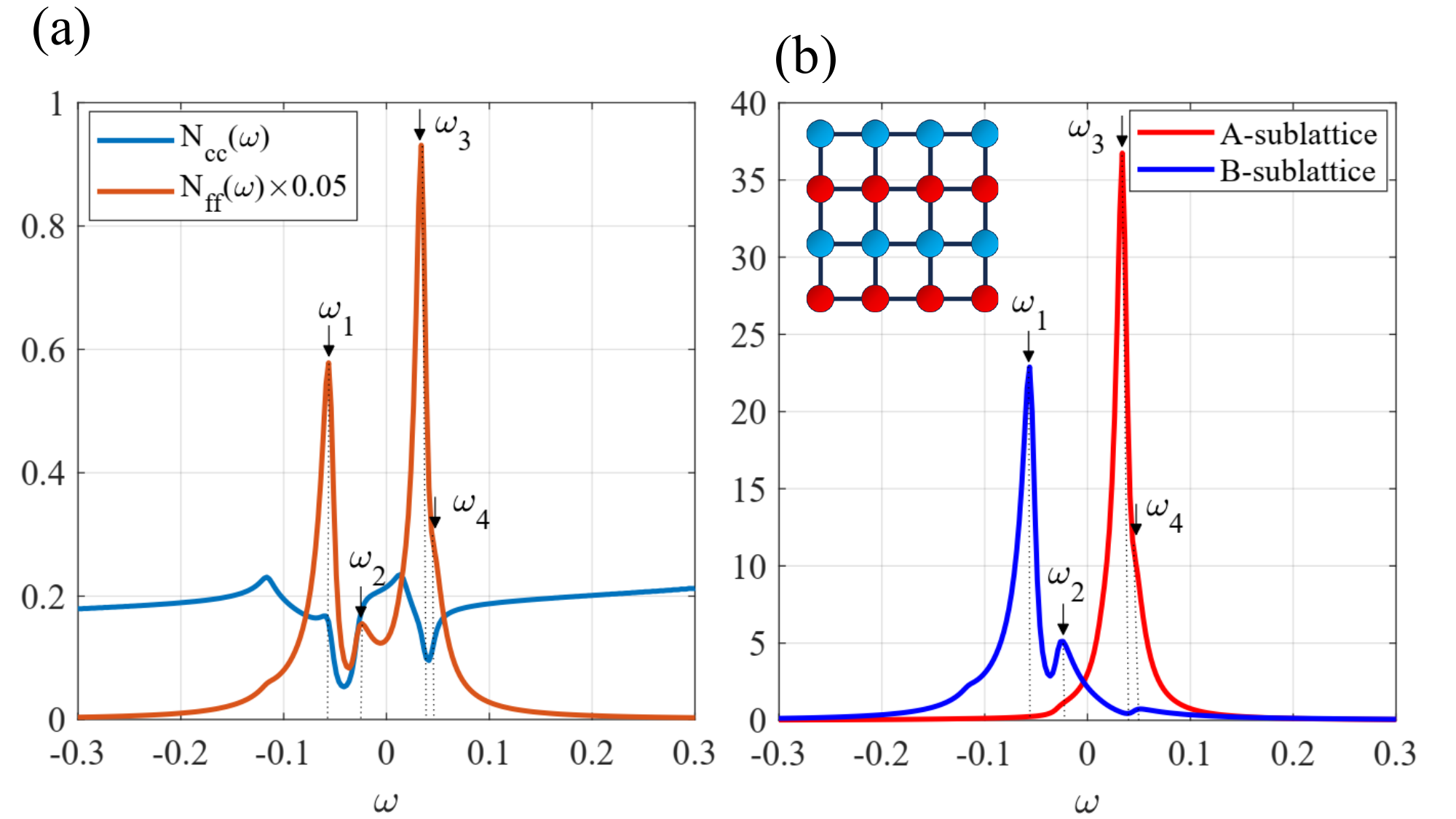}
\caption{\label{fig:4} (a) The DOS of conduction electron $N_{cc}(\omega)$ and Abrikosov fermion $N_{ff}(\omega)$ and (b) LDOS of Abrikosov fermion on A and B sublattice for KHW phase with $\boldsymbol{Q}=(0,\pi)$.}
\end{figure}

\emph{The KHW with $\boldsymbol{Q}=(0,\pi)$}.--
We now focus on the canonical stripe-type KHW phase at $\boldsymbol{Q}=(0,\pi)$, stabilized at low electron density $n_c=0.25$ and intermediate Kondo coupling $J=2$. The quasiparticle band structure, density of states (DOS), and real-space local density of states (LDOS) for this phase are summarized in Figs.~\ref{fig:3} and \ref{fig:4}. Consistent with our lattice construction, the spatially modulated hybridization ansatz $V_{i}=V_{0}+V_{1}e^{i\boldsymbol{Q}\cdot\boldsymbol{R}_{i}}$ induces a uniaxial stripe modulation and doubles the unit cell along the $y$ direction (Fig.~\ref{fig:1}(b)). Consequently, the resulting KHW Hamiltonian yields four distinct quasiparticle bands (Fig.~\ref{fig:3}), in stark contrast to the two hybridized bands obtained in the conventional spatially uniform Kondo state.
These emergent bands retain strong $f$-$c$ hybridization and exhibit two characteristic energy gaps: a dominant primary gap $\Delta_{0}\approx0.023$ and a weaker secondary gap $\Delta_{1}\approx0.01$ (Fig.~\ref{fig:3}(b)). These gaps can be quantitatively understood within standard Kondo hybridization theory\cite{Coleman2015}. Using the self-consistent solutions $V_{0}=0.135$ and $V_{1}=0.049$, the analytical estimations $\Delta_{0}=(JV_{0})^{2}/4t\approx0.018$ and $\Delta_{1}=(JV_{1})^{2}/4t\approx0.002$ agree well with our numerical results. This confirms that both gaps originate intrinsically from modulated Kondo hybridization, sharing the same microscopic mechanism as uniform heavy-fermion hybridization gaps.

The Abrikosov fermion DOS, defined as $N_{ff}(\omega)=N_s^{-1}\sum_{k}A_{ff}(k,\omega)$ with $A_{ff}(k,\omega)$ the fermion spectral function, is displayed in Fig.~\ref{fig:4}(a). The prominent spectral peaks at $\omega_1$, $\omega_2$, and $\omega_3$ directly correspond to the flat-band manifolds in the quasiparticle spectrum. The main hybridization gap $\Delta_{0}$ is precisely captured by the energy separation $\omega_2-\omega_1$. In contrast, the smaller secondary gap $\Delta_1$ cannot be fully resolved from the primary spectral features due to its suppressed magnitude. Nevertheless, a weak spectral kink emerging at $\omega_4$ unambiguously defines the secondary gap scale $\Delta_1=\omega_4-\omega_3$. The corresponding conduction-electron spectral properties follow similar behavior.

Owing to the $y$-direction unit-cell doubling induced by $\boldsymbol{Q}=(0,\pi)$ order, the system decomposes into inequivalent $A$ and $B$ sublattices with strongly distinct LDOS responses (Fig.~\ref{fig:4}(b)). Specifically, the $A$-sublattice LDOS is governed by the secondary gap $\Delta_1$, while the $B$-sublattice spectrum features the dominant hybridization gap $\Delta_0$. Adopting the experimental interpretation that the STM $dI/dV$ signal is dominated by Abrikosov fermion contributions\cite{Cao2026}, our LDOS results reproduce the key experimental modulation pattern. The extracted hybridization gap remains homogeneous along the $x$ direction but oscillates periodically between $\Delta_1$ and $\Delta_0$ along the $y$ direction across alternating sublattice sites. This spontaneous uniaxial unit-cell-doubling modulation of the hybridization gap provides a direct microscopic explanation for the STM spectroscopic signatures observed in 1T/1H-TaS$_2$\cite{Cao2026}.


\begin{figure*}[!t]
\includegraphics[width=0.5\linewidth]{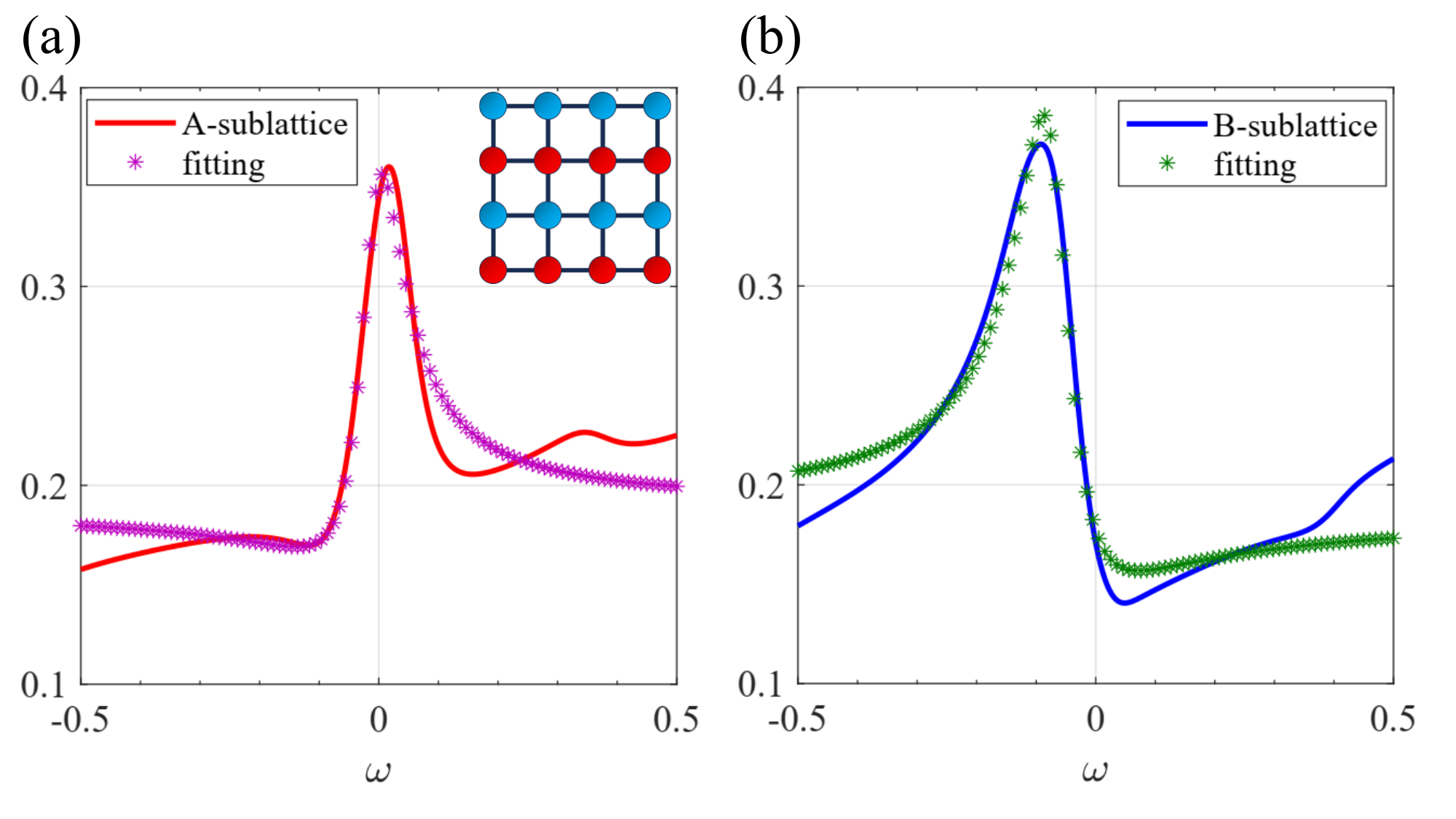}
\includegraphics[width=0.5\linewidth]{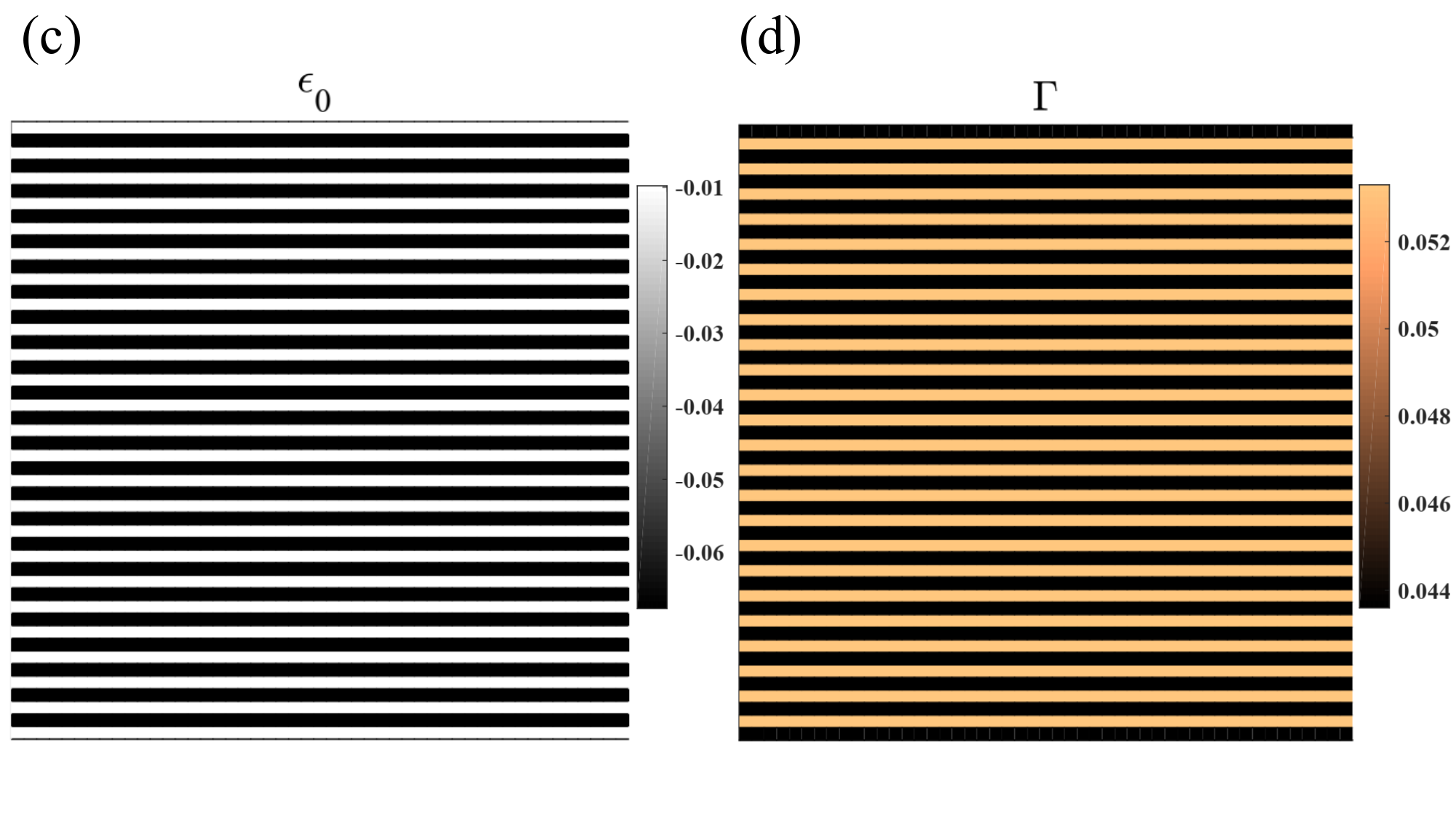}
\caption{\label{fig:11}(a)(b) The calculated STM spectrum and its fitting for $A$ and $B$ sublattice. (c)(d) The stripe-like spatial distribution of fitting parameters $\epsilon_{0}$ and $\Gamma$ for KHW with $\boldsymbol{Q}=(0,\pi)$.}
\end{figure*}
\emph{Relevance and implication to experiments}.--
In the spin-triplet superconducting candidate UTe$_2$, Fano-line-shape fittings of STM data uncover pronounced stripe-like spatial modulations of the Kondo hybridization strength\cite{Yu2026}, which precisely matches the characteristic spatial pattern of our theoretically predicted $\boldsymbol{Q}=(0,\pi)$ KHW phase as shown in Fig.~\ref{fig:11}. Specifically, the calculated STM spectrum on two sublattices illustrated in Figs.~\ref{fig:11}(a) and (b) is fitted by Fano line-shape formula $\frac{(q+(\omega-\epsilon_{0})/\Gamma)^2}{1+((\omega-\epsilon_{0})/\Gamma)^2}$ with fitting parameters $q$, $\epsilon_{0}$ and $\Gamma$\cite{Schmidt2010,Yu2026}. The spatial distribution of fitting parameters $\epsilon_{0}$ and $\Gamma$ are illustrated in Figs.~\ref{fig:11}(c) and (d), which reveals the identical stripe structure observed in Ref.~\onlinecite{Yu2026}. (see SM for details on STM spectrum and its fitting) Crucially, our theory realizes such modulated hybridization order in the absence of conventional charge-density-wave instabilities. This key feature demonstrates that KHW ordering can emerge purely from spontaneous modulation of the Kondo hybridization field, establishing it as an independent electronic order distinct from conventional charge ordering.

\begin{figure}
\includegraphics[width=0.95\linewidth]{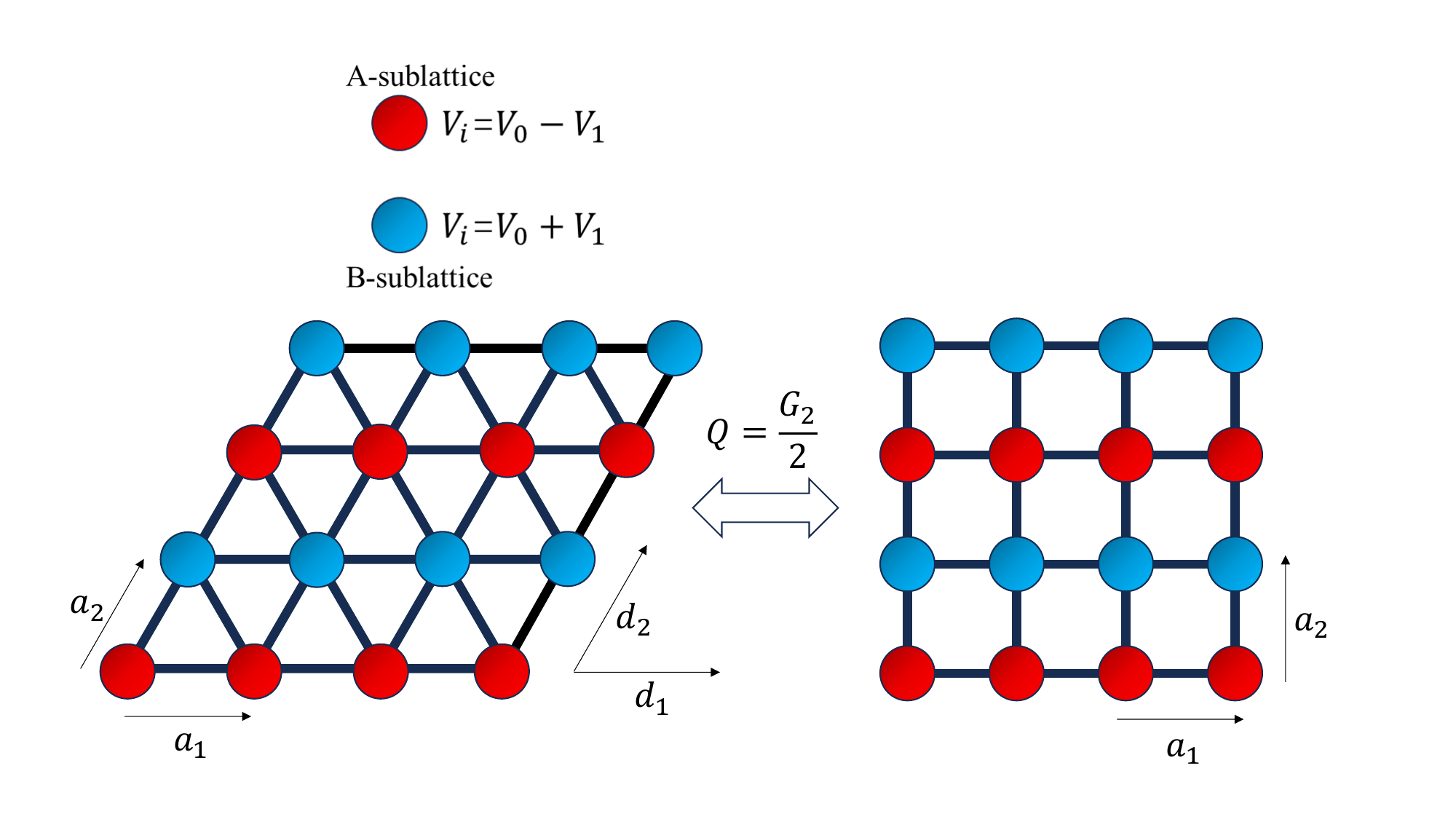}
\caption{\label{fig:5} The stripe-like structure of Kondo hybridization order parameter $V_{i}=V_{0}+V_{1}e^{i\boldsymbol{Q}\cdot \boldsymbol{R_{i}}}$ with $\boldsymbol{Q}=\frac{\boldsymbol{G_{2}}}{2}$ on triangular and square lattice.}
\end{figure}

Experimental evidence for KHW physics also arises from STM measurements on the artificial triangular Kondo lattice/periodic Anderson model system 1T/1H-TaS$_2$\cite{Chen2021,Cao2026}. Spatially resolved spectroscopic results reveal anisotropic oscillatory modulations of the hybridization gap: the modulation retains the intrinsic lattice period along one high-symmetry direction $d_1$, while exhibiting a characteristic unit-cell-doubling periodicity along the orthogonal direction $d_2$. When mapped to the effective square-lattice geometry, this directional doubling behavior exactly corresponds to the $\boldsymbol{Q}=(0,\pi)$ KHW order obtained in our calculations, providing solid theoretical interpretation for the experimental observations in 1T/1H-TaS$_2$.
Quantitatively, the STM data for 1T/1H-TaS$_2$ supports a spatially modulated hybridization ansatz $V_{i}=V_{0}+V_{1}e^{i\boldsymbol{Q}\cdot\boldsymbol{R}_{i}}$ defined on the triangular lattice, with the ordering wavevector taking the universal form $\boldsymbol{Q}=\boldsymbol{G}_{2}/2$. Here, $\boldsymbol{G}_{1}=2\pi(1,-1/\sqrt{3})$ and $\boldsymbol{G}_{2}=2\pi(0,2/\sqrt{3})$ denote the primitive reciprocal lattice vectors of the triangular lattice. The resultant KHW state generates uniaxial stripe modulation along the $d_1$ direction and unit-cell doubling along the $d_2$ direction, achieving full consistency with experimental STM spectra (see Fig.~\ref{fig:5}).
For square-lattice Kondo systems, the reciprocal lattice vectors are $\boldsymbol{G}_{1}=(2\pi,0)$ and $\boldsymbol{G}_{2}=(0,2\pi)$, and the canonical stripe-type KHW wavevector $\boldsymbol{Q}=(0,\pi)$ likewise satisfies $\boldsymbol{Q}=\boldsymbol{G}_{2}/2$. This unified wavevector construction holds for both square and triangular lattices, revealing a universal organizing principle for stripe-like KHW order across different lattice geometries. This universal scenario validates the generalization of our square-lattice KHW theory to triangular Kondo lattice platforms, demonstrating the broad applicability of our microscopic framework.

To further characterize the symmetry-broken nature of the KHW state, we investigate its zero-temperature transport response via linear-response theory. The conductivity tensor is given by $\sigma^{\alpha\beta}=\frac{\pi e^{2}}{N_{s}}\sum_{k}\frac{\partial \varepsilon_{k}}{\partial k^{\alpha}}\frac{\partial \varepsilon_{k}}{\partial k^{\beta}}(A_{cc}(k,0))^{2}$ where $A_{cc}(\boldsymbol{k},0)$ is the zero-frequency spectral function of itinerant conduction electrons\cite{Coleman2015,Mahan2000}. For the stripe-ordered KHW phase at $\boldsymbol{Q}=(0,\pi)$, the system retains translational invariance along the $x$ direction while developing a two-sublattice periodic modulation along the $y$ axis. This uniaxial modulation spontaneously breaks the in-plane $x$-$y$ rotational symmetry, giving rise to inherently anisotropic electrical transport with $\sigma^{xx}\neq\sigma^{yy}$. We confirm this scenario through self-consistent numerical calculations at $J=2$ and $n_c=0.25$, which yield distinct conductivity components $\sigma^{xx}=2.6568$ and $\sigma^{yy}=3.7316$. (electron charge $e=1$ and a damping factor $\Gamma=0.001$ are utilized) The clear discrepancy directly evidences symmetry lowering induced by the stripe-type KHW ordering.

In contrast, the checkerboard KHW state at $\boldsymbol{Q}=(\pi,\pi)$ preserves full in-plane rotational symmetry, enforcing isotropic transport behavior $\sigma^{xx}=\sigma^{yy}$. Our calculations at $J=4$ and $n_c=1$ yield identical diagonal conductivities $\sigma^{xx}=\sigma^{yy}=3.6864\times 10^{-4}$. The strongly suppressed magnitude reflects the intrinsic insulating band structure of the $\boldsymbol{Q}=(\pi,\pi)$ phase. Likewise, the uniform Kondo hybridization state at $\boldsymbol{Q}=(0,0)$ maintains unbroken rotational symmetry and exhibits isotropic conductivity. These results establish in-plane conductivity anisotropy as a robust, experimentally feasible fingerprint for the symmetry-broken $\boldsymbol{Q}=(0,\pi)$ stripe KHW phase. Such transport signatures enable clear discrimination between uniaxial stripe order, uniform Kondo backgrounds, and symmetric checkerboard KHW states. We propose that anisotropic transport measurements along orthogonal crystalline axes can serve as a powerful experimental probe to identify and characterize emergent KHW ordering in heavy-fermion systems, including UTe$_2$ and 1T/1H-TaS$_2$.

\emph{Conclusion}.--
We have established the microscopic existence of spatially modulated Kondo hybridization wave (KHW) states in the square-lattice Kondo lattice model. Despite adopting the large-$N$ fermionic mean-field approximation, our minimal theoretical framework successfully captures the essential structural and spectroscopic signatures of emergent KHW order. The obtained stripe-type modulations and unit-cell-doubling structures are in excellent consistency with recent STM observations on UTe$_2$ and 1T/1H-TaS$_2$\cite{Yu2026,Cao2026}, offering a transparent microscopic basis for further explorations on the exotic intertwined electronic instabilities and nematic-like hybridization phenomena in these systems. While mean-field theory provides a qualitatively reliable description of KHW phase, sophisticated numerical techniques, including neural-quantum-state variational Monte Carlo and tensor-network simulations, can incorporate intrinsic quantum fluctuations and promise more precise benchmarking of KHW phase diagrams and transport properties for future studies\cite{Nikolaenko2026,Chen2022}.

\emph{Note added}.--After we post our paper on arXiv, we note a stripe channel order has been found in $SU(N)$ two-channel Kondo lattice model\cite{Cohen2026}, with the help of determinant quantum Monte Carlo simulations and large-$N$ saddle point analysis.

\section*{Acknowledgments}
We thank Hui Yin for preparing the phase diagram, Qianqian Shi and Wei-Wei Yang for their careful reading and collaboration on related issues. This work was supported by the National Natural Science
Foundation of China (Grant No. 12247101), the Fundamental Research Funds for the Central Universities (Grant
No. lzujbky-2024-jdzx06), the Natural Science Foundation
of Gansu Province (Grants No. 22JR5RA389 and No. 25JRRA799), and the National ''111 Center'' for Collaborative Research (Grant No. B20063).
\section*{DATA AVAILABILITY}
The data that support the findings of this article are not
publicly available upon publication because it is not technically feasible and/or the cost of preparing, depositing, and hosting the data would be prohibitive within the terms of this
research project. The data are available from the authors upon
reasonable request.

\newpage
\section*{Supplementary Materials}
\subsection{Mean-field Hamiltonian and self-consistent equations}
We start with the standard Kondo lattice model (Eq.~\ref{eq1})
\begin{equation}
H=\sum_{k\sigma}\varepsilon_{k}c_{k\sigma}^{\dag}c_{k\sigma}+J\sum_{i}\vec{S}_{i}^{c}\cdot\vec{S}_{i}^{f},
\end{equation}
which describes the interplay between conduction electron $c_{k\sigma}$ and local moment of $f$-electron $\vec{S}_{i}^{f}$. This model can be defined on any regular lattice and we here consider it is defined on the square lattice. The fermionic operator $c_{k\sigma}$ satisfies the usual anti-commutative relations $[c_{k\sigma},c_{k'\sigma'}^{\dag}]_{+}=\delta_{k,k'}\delta_{\sigma,\sigma'}$ and $[c_{k\sigma},c_{k'\sigma'}]_{+}=0$.
$\vec{S}_{i}^{f}=\frac{\vec{\sigma}^{f}_{i}}{2}$ denotes the spin degree of freedom for active $f$-electron in heavy fermion compounds with $4f/5f$ elements. On square lattice, the bare dispersion of conduction electron is $\varepsilon_{k}=-2t(\cos k_{x}+\cos k_{y})-4t'\cos k_{x}\cos k_{y}-\mu$, with nearest-neighbor-hopping $t$, next-nearest-neighbor-hopping $t'$ and chemical potential $\mu$. The Kondo lattice model presented above is hard to solve on two-dimensional lattice if the density of conduction electron is deviated from the half-filling case $n_{c}=1$. (The half-filled Kondo lattice model on square lattice has been solved by quantum Monte Carlo simulation and ones find insulating antiferromagnetic phase and Kondo insulator\cite{Assaad1999}.)

To proceed, slave-particle methods are widely used to capture the basic feature of the mentioned Kondo lattice model. Here, we utilize Abrikosov fermion representation for $\vec{S}_{i}^{f}$,
\begin{equation}
\vec{S}_{i}^{f}=\frac{1}{2}\sum_{\sigma\sigma'}f^{\dag}_{i\sigma}\vec{\sigma}_{\sigma\sigma'}f_{i\sigma'},
\end{equation}
which splits $S=1/2$-spin into two fermions $f_{i\uparrow}$ and $f_{i\downarrow}$. These two kinds of fermions do not have charge degree of freedom and they can be considered as spinon in the language of quantum spin liquid. At the same time, there exists the single-occupancy constraint $\sum_{\sigma}f_{i\sigma}^{\dag}f_{i\sigma}=1$, which should be imposed on each lattice site. Therefore, we have
\begin{equation}
\vec{S}_{i}^{c}\cdot\vec{S}_{i}^{f}=-\frac{1}{2}\sum_{\sigma\sigma'}c_{i\sigma}^{\dag}f_{i\sigma}f_{i\sigma'}^{\dag}c_{i\sigma'}+\frac{1}{4}\sum_{\sigma} c_{i\sigma}^{\dag}c_{i\sigma}.
\end{equation}
The latter term $\frac{1}{4}\sum_{\sigma} c_{i\sigma}^{\dag}c_{i\sigma}$ can be absorbed into the redefinition of chemical potential, i.e. $\mu\rightarrow\mu-\frac{J}{4}$. Now, we can write down the corresponding path integral formalism,
\begin{equation}
\mathcal{Z}=\int \mathcal{D}\bar{c}\mathcal{D}c\mathcal{D}\bar{f}\mathcal{D}f\mathcal{D}\lambda e^{-S}\nonumber
\end{equation}
\begin{widetext}
\begin{equation}
S=\int_{0}^{\beta}d\tau\left[\sum_{k\sigma}\bar{c}_{k\sigma}(\partial_{\tau}+\varepsilon_{k})c_{k\sigma}+\sum_{i\sigma}\bar{f}_{i\sigma}(\partial_{\tau}+i\lambda_{i})f_{i\sigma}
-\frac{J}{2}\sum_{i,\sigma\sigma'}\bar{c}_{i\sigma}f_{i\sigma}\bar{f}_{i\sigma'}c_{i\sigma'}
-i\sum_{i}\lambda_{i}\right].\nonumber
\end{equation}
\end{widetext}
Here, $\mathcal{Z}=\mathrm{Tr}e^{-\beta H}$ is the partition function and $S$ denotes the action in the imaginary-time ($\tau\in[0,\beta]$) formalism. Note that $\lambda_{i}(\tau)$ is the dynamic Lagrange multiplier, which has both spatial and time dependence. Usually, we assume
$\lambda_{i}$ is static and neglect its time dependence with redefinition $i\lambda_{i}\rightarrow\lambda_{i}$. The interaction term $\bar{c}_{i\sigma}f_{i\sigma}\bar{f}_{i\sigma'}c_{i\sigma'}$ can be decoupled with the help of Hubbard-Stratonovich transformation.\cite{Coleman2015} If the introduced bosonic field in Hubbard-Stratonovich transformation is considered to be static, the treatment here is equivalent to the mean-field decoupling in the Hamiltonian formalism:
\begin{widetext}
\begin{equation}
c_{i\sigma}^{\dag}f_{i\sigma}f_{i\sigma'}^{\dag}c_{i\sigma'}
\simeq
\langle c_{i\sigma}^{\dag}f_{i\sigma}\rangle f_{i\sigma'}^{\dag}c_{i\sigma'}
+c_{i\sigma}^{\dag}f_{i\sigma}\langle f_{i\sigma'}^{\dag}c_{i\sigma'}\rangle
-\langle c_{i\sigma}^{\dag}f_{i\sigma}\rangle\langle f_{i\sigma'}^{\dag}c_{i\sigma'}\rangle,
\end{equation}
\end{widetext}
where we introduce the Kondo hybridization/screening order parameter as
\begin{equation}
V_{i}^{\sigma}=\langle c_{i\sigma}^{\dag}f_{i\sigma}\rangle,
\end{equation}
although the involved $U(1)$ gauge symmetry cannot be broken. The nonzero value of $V_{i}^{\sigma}$ denotes the appearance of Kondo hybridization in the mean-field level. Now, the Kondo interaction term is rewritten as
\begin{widetext}
\begin{equation}
H_{K}=J\sum_{i}\vec{S}_{i}^{c}\cdot\vec{S}_{i}^{f}
\simeq-\frac{J}{2}\sum_{i,\sigma\sigma'}[V_{i}^{\sigma}f_{i\sigma'}^{\dag}c_{i\sigma'}
+(V_{i}^{\sigma'})^{\ast}c_{i\sigma}^{\dag}f_{i\sigma}
-V_{i}^{\sigma}(V_{i}^{\sigma'})^{\ast}].
\end{equation}
\end{widetext}
Meanwhile, the constraint part in the path integral formalism gives the following term
\begin{eqnarray}
H_{\lambda}=\sum_{i}\lambda_{i}\left(\sum_{\sigma}f_{i\sigma}^{\dag}f_{i\sigma}-1\right).
\end{eqnarray}
So, the mean-field Hamiltonian for Kondo lattice model reads as
\begin{widetext}
\begin{equation}
H=\sum_{k\sigma}\varepsilon_{k}c_{k\sigma}^{\dag}c_{k\sigma}-\frac{J}{2}\sum_{i,\sigma\sigma'}[V_{i}^{\sigma}f_{i\sigma'}^{\dag}c_{i\sigma'}
+(V_{i}^{\sigma'})^{\ast}c_{i\sigma}^{\dag}f_{i\sigma}
-V_{i}^{\sigma}(V_{i}^{\sigma'})^{\ast}]+\sum_{i}\lambda_{i}\left(\sum_{\sigma}f_{i\sigma}^{\dag}f_{i\sigma}-1\right).
\end{equation}
\end{widetext}
In conventional treatment, $V_{i}^{\sigma}$ is assumed to be spatially uniform and have no spin dependence if only paramagnetic states are considered. Such simplification leads to $V_{i}^{\sigma}=V$ but it cannot explain the Kondo hybridization wave state observed in experiments. Thus, we are forced to relax the assumption on $V_{i}^{\sigma}$. It is found that, the following ansatz works
\begin{equation}
V_{i}^{\sigma}=V_{i}=V_{0}+V_{1}e^{i\boldsymbol{Q}\cdot\boldsymbol{R_{i}}},
\end{equation}
where $V_{0},V_{1}$ represent the amplitude of the uniform and oscillating part for Kondo hybridization order parameter. They will be chosen to be real to simplify our treatment. $\boldsymbol{Q}$ is the characteristic wavevector and $\boldsymbol{R_{i}}$ is the position vector of lattice $i$. When $V_{1},\boldsymbol{Q}\neq0$, one finds the Kondo hybridization wave state. Inserting the ansatz  $V_{i}=V_{0}+V_{1}e^{i\boldsymbol{Q}\cdot\boldsymbol{R_{i}}}$ into $H_{K}$,
\begin{widetext}
\begin{eqnarray}
H_{K}&=&J\sum_{i}\vec{S}_{i}^{c}\cdot \vec{S}_{i}^{f}=-\frac{J}{2}\sum_{i,\sigma\sigma'}\left[V_{i}^{\sigma}f_{i\sigma'}^{\dag}c_{i\sigma'}+(V_{i}^{\sigma})^{\ast}c_{i\sigma'}^{\dag}f_{i\sigma'} -V_{i}^{\sigma}(V_{i}^{\sigma})^{\ast}\right]\nonumber\\
&=&-\frac{J}{2}\sum_{i,\sigma\sigma'}\left[(V_{0}+V_{1}e^{i\boldsymbol{Q}\cdot\boldsymbol{R_{i}}})f_{i\sigma'}^{\dag}c_{i\sigma'}+(V_{0}+V_{1}e^{-i\boldsymbol{Q}\cdot\boldsymbol{R_{i}}})c_{i\sigma'}^{\dag}f_{i\sigma'} -(V_{0}+V_{1}e^{i\boldsymbol{Q}\cdot\boldsymbol{R_{i}}})(V_{0}+V_{1}e^{-i\boldsymbol{Q}\cdot\boldsymbol{R_{i}}})\right]\nonumber\\
&=&-JV_{0}\sum_{i,\sigma}\left[f_{i\sigma}^{\dag}c_{i\sigma}+c_{i\sigma}^{\dag}f_{i\sigma}\right]
-JV_{1}\sum_{i,\sigma}\left[e^{i\boldsymbol{Q}\cdot\boldsymbol{R_{i}}}f_{i\sigma}^{\dag}c_{i\sigma}+e^{-i\boldsymbol{Q}\cdot\boldsymbol{R_{i}}}c_{i\sigma}^{\dag}f_{i\sigma}\right]
+2JN_{s}(V_{0}^{2}+V_{1}^{2}+2V_{0}V_{1}\delta_{Q,0})\nonumber\\
&=&-JV_{0}\sum_{k\sigma}\left[f_{k\sigma}^{\dag}c_{k\sigma}+c_{k\sigma}^{\dag}f_{k\sigma}\right]
-JV_{1}\sum_{k\sigma}\left[f_{k+Q\sigma}^{\dag}c_{k\sigma}+c_{k\sigma}^{\dag}f_{k+Q\sigma}\right]
+2JN_{s}(V_{0}^{2}+V_{1}^{2}+2V_{0}V_{1}\delta_{Q,0}).
\end{eqnarray}
\end{widetext}
We see that the conduction electron with momentum $k$ couples to Abrikosov fermion with momentum $k$ and $k+Q$. In contrast, the usual treatment only involves the coupling between the ones with the same momentum $k$.

Since Kondo hybridization order parameter has spatial dependence, the Lagrange multiplier $\lambda_{i}$ is also nonuniform and has the same $\boldsymbol{Q}$-dependence as $V_{i}$, i.e.
\begin{equation}
\lambda_{i}=\lambda\frac{e^{i\boldsymbol{Q}\cdot \boldsymbol{R_{i}}}+e^{-i\boldsymbol{Q}\cdot \boldsymbol{R_{i}}}}{2}
=\lambda\cos(\boldsymbol{Q}\cdot \boldsymbol{R_{i}}),
\end{equation}
so the Lagrange multiplier term is written as
\begin{widetext}
\begin{eqnarray}
H_{\lambda}&=&\sum_{i}\lambda_{i}\left(\sum_{\sigma}f_{i\sigma}^{\dag}f_{i\sigma}-1\right)=\frac{\lambda}{2}\sum_{i\sigma}(e^{i\boldsymbol{Q}\cdot\boldsymbol{R_{i}}}+e^{-i\boldsymbol{Q}\cdot\boldsymbol{R_{i}}})f_{i\sigma}^{\dag}f_{i\sigma}
-\lambda\sum_{i}\cos(\boldsymbol{Q}\cdot\boldsymbol{R_{i}})\nonumber\\
&=&\frac{\lambda}{2}\sum_{k\sigma}(f_{k\sigma}^{\dag}f_{k+Q\sigma}+f_{k+Q\sigma}^{\dag}f_{k\sigma})-N_{s}\lambda\delta_{Q=0}.
\end{eqnarray}
\end{widetext}
In this case, Abrikosov fermions with momentum $k$ and $k+Q$ are coupled, which is different to the spatially uniform situation $\lambda\sum_{k\sigma}f_{k\sigma}^{\dag}f_{k\sigma}$. Adding $H_{K}$ and $H_{\lambda}$ into the mean-field Hamiltonian $H$, we have
\begin{widetext}
\begin{eqnarray}
H&=&\sum_{k\sigma}\varepsilon_{k}c_{k\sigma}^{\dag}c_{k\sigma}+\frac{\lambda}{2}\sum_{k\sigma}(f_{k\sigma}^{\dag}f_{k+Q\sigma}+f_{k+Q\sigma}^{\dag}f_{k\sigma})-JV_{0}\sum_{k\sigma}\left[f_{k\sigma}^{\dag}c_{k\sigma}+c_{k\sigma}^{\dag}f_{k\sigma}\right]
-JV_{1}\sum_{k\sigma}\left[f_{k+Q\sigma}^{\dag}c_{k\sigma}+c_{k\sigma}^{\dag}f_{k+Q\sigma}\right]\nonumber\\
&+&N_{s}\left[2J(V_{0}^{2}+V_{1}^{2}+2V_{0}V_{1}\delta_{Q=0})-\lambda\delta_{Q=0}\right]\nonumber\\
&=&\frac{1}{2}\sum_{k\sigma}\left(
                   \begin{array}{cccc}
                     c_{k\sigma}^{\dag} & f_{k+Q\sigma}^{\dag} & c_{k+Q\sigma}^{\dag} & f_{k\sigma}^{\dag} \\
                   \end{array}
                 \right)\left(
                          \begin{array}{cccc}
                            \varepsilon_{k} & -JV_{1} & 0 & -JV_{0} \\
                            -JV_{1} & 0 & -JV_{0} & \lambda \\
                            0 & -JV_{0} & \varepsilon_{k+Q} & -JV_{1} \\
                            -JV_{0} & \lambda & -JV_{1} & 0 \\
                          \end{array}
                        \right)\left(
                                 \begin{array}{c}
                                   c_{k\sigma} \\
                                   f_{k+Q\sigma} \\
                                   c_{k+Q\sigma} \\
                                   f_{k\sigma} \\
                                 \end{array}
                               \right)\nonumber\\
&+&N_{s}\left[2J(V_{0}^{2}+V_{1}^{2}+2V_{0}V_{1}\delta_{Q=0})-\lambda\delta_{Q=0}\right].
\end{eqnarray}
\end{widetext}
If we introduce the spinor $\Psi_{k\sigma}^{\dag}=(c_{k\sigma}^{\dag},f_{k+Q\sigma}^{\dag},c_{k+Q\sigma}^{\dag},f_{k\sigma}^{\dag})$ and Bloch Hamiltonian
\begin{widetext}
\begin{equation}
H_{k}=\frac{1}{2}\left(
                          \begin{array}{cccc}
                            \varepsilon_{k} & -JV_{1} & 0 & -JV_{0} \\
                            -JV_{1} & 0 & -JV_{0} & \lambda \\
                            0 & -JV_{0} & \varepsilon_{k+Q} & -JV_{1} \\
                            -JV_{0} & \lambda & -JV_{1} & 0 \\
                          \end{array}
                        \right),
\end{equation}
\end{widetext}
we can obtain the one used in the main text,
\begin{widetext}
\begin{equation}
H=\sum_{k\sigma}\Psi_{k\sigma}^{\dag}H_{k}\Psi_{k\sigma}+N_{s}\left[2J(V_{0}^{2}+V_{1}^{2}+2V_{0}V_{1}\delta_{Q=0})-\lambda\delta_{Q=0}\right].\nonumber
\end{equation}
\end{widetext}
In addition to the mean-field Hamiltonian, there are three kinds of equations to be solved,
\begin{equation}
V_{i}=V_{0}+V_{1}e^{i\boldsymbol{Q}\cdot\boldsymbol{R_{i}}}=\langle c_{i\sigma}^{\dag}f_{i\sigma}\rangle,\nonumber
\end{equation}
\begin{equation}
1=\sum_{\sigma}\langle f_{i\sigma}^{\dag}f_{i\sigma}\rangle,\nonumber
\end{equation}
\begin{equation}
n_{c}=\frac{1}{N_{s}}\sum_{i\sigma}\langle c_{i\sigma}^{\dag}c_{i\sigma}\rangle.\nonumber
\end{equation}
Transform them into momentum space, we have
\begin{equation}
V_{0}=\frac{1}{N_{s}}\sum_{k}\langle c_{k\sigma}^{\dag}f_{k\sigma}\rangle,
\end{equation}
\begin{equation}
V_{1}=\frac{1}{N_{s}}\sum_{k}\langle c_{k\sigma}^{\dag}f_{k+Q\sigma}\rangle,
\end{equation}
\begin{equation}
1=\frac{1}{N_{s}}\sum_{k\sigma}\langle f_{k\sigma}^{\dag}f_{k\sigma}\rangle,
\end{equation}
and
\begin{equation}
n_{c}=\frac{1}{N_{s}}\sum_{k\sigma}\langle c_{k\sigma}^{\dag}c_{k\sigma}\rangle.
\end{equation}
Here, we show how to obtain the first two equations for $V_{0},V_{1}$. Use the Fourier transformation $c_{i\sigma}=\frac{1}{\sqrt{N_{s}}}\sum_{k}e^{ikR_{i}}c_{k\sigma}$, we obtain
\begin{eqnarray}
\langle c_{i\sigma}^{\dag}f_{i\sigma}\rangle=\frac{1}{N_{s}}\sum_{k,k'}e^{-ikR_{i}}e^{ik'R_{i}}\langle c_{k\sigma}^{\dag}f_{k'\sigma}\rangle.\nonumber
\end{eqnarray}
Since the Hamiltonian couples $c_{k\sigma}^{\dag}$ with $f_{k+Q\sigma}$ ($f_{k\sigma}$) via $-JV_{1}$ ($-JV_{0}$), we must have
\begin{equation}
\langle c_{k\sigma}^{\dag}f_{k'\sigma}\rangle=\delta_{k',k}\langle c_{k\sigma}^{\dag}f_{k\sigma}\rangle+\delta_{k',k+Q}\langle c_{k\sigma}^{\dag}f_{k+Q\sigma}\rangle,\nonumber
\end{equation}
thus
\begin{eqnarray}
\langle c_{i\sigma}^{\dag}f_{i\sigma}\rangle=\frac{1}{N_{s}}\sum_{k}\langle c_{k\sigma}^{\dag}f_{k\sigma}\rangle+e^{i\boldsymbol{Q}\cdot\boldsymbol{R_{i}}}\frac{1}{N_{s}}\sum_{k}\langle c_{k\sigma}^{\dag}f_{k+Q\sigma}\rangle.\nonumber
\end{eqnarray}
Finally, compare it with $V_{0}+V_{1}e^{i\boldsymbol{Q}\cdot\boldsymbol{R_{i}}}$, we obtain the desirable equations $V_{0}=\frac{1}{N_{s}}\sum_{k}\langle c_{k\sigma}^{\dag}f_{k\sigma}\rangle$ and $V_{1}=\frac{1}{N_{s}}\sum_{k}\langle c_{k\sigma}^{\dag}f_{k+Q\sigma}\rangle$.

Before ending this section, it is noted that when $\boldsymbol{Q}=(0,0)$ (the case for the spatially uniform Kondo hybridization), the equations for $V_{0},V_{1}$ reduce into the same one, which means the present mean-field formalism will be singular if $\boldsymbol{Q}\rightarrow(0,0)$. Therefore, we will not use the above mean-field equations for studying $\boldsymbol{Q}=(0,0)$ states, and the more useful and exact formalism will be given in later part of this SM.
\subsection{Ground-state phase diagram}

\begin{figure}
\includegraphics[width=0.95\linewidth]{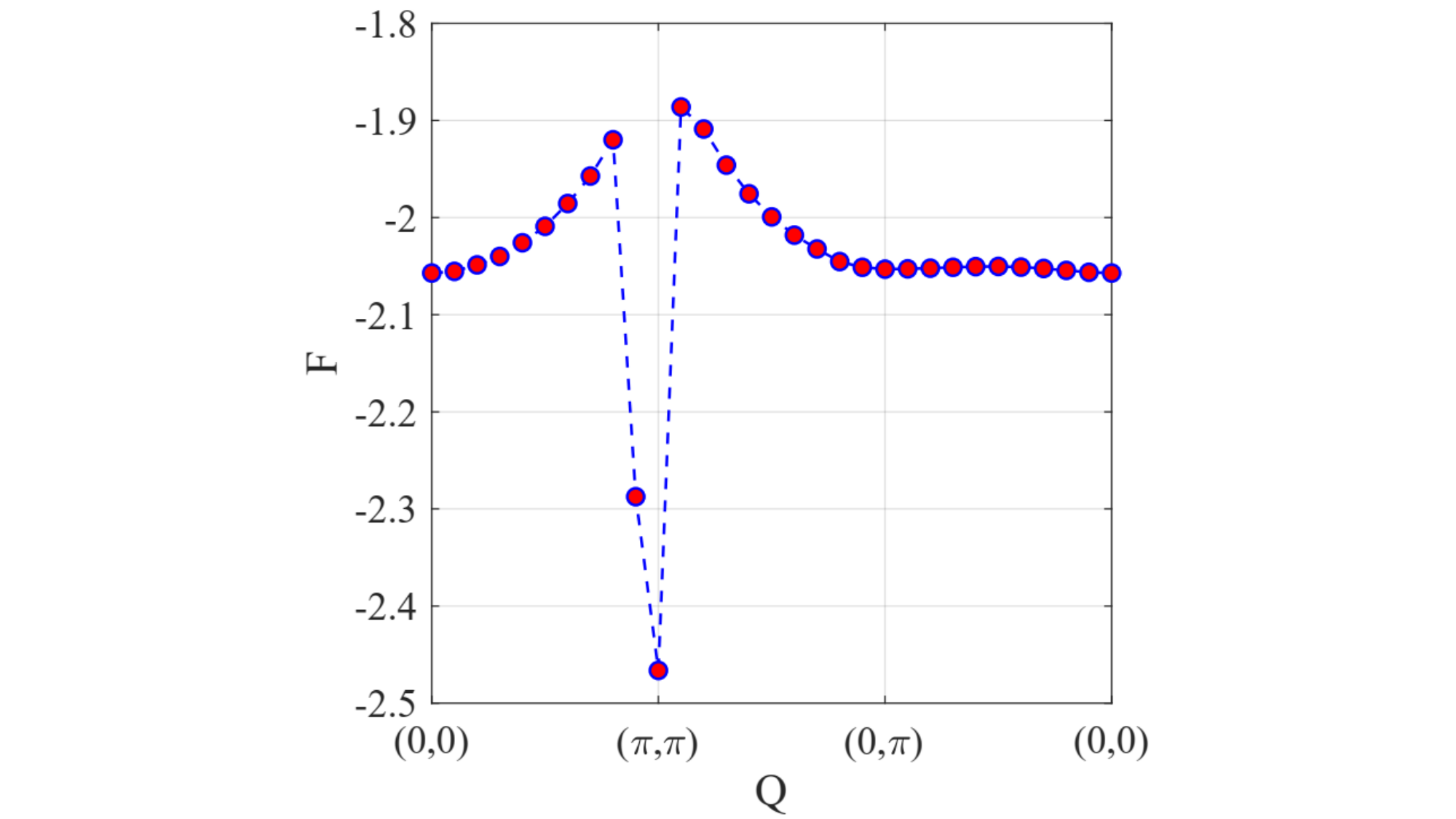}
\caption{\label{fig:6} The free energy density $F$ versus $\boldsymbol{Q}$ for half-filling case with $J=4$.}
\end{figure}

We have shown some typical data used for constructing the ground-state phase diagram Fig.~\ref{fig:2} in Table~\ref{table_1}. These data are obtained as follows. Firstly, we choose fixed Kondo interaction $J$ and conduction electron density $n_{c}$. Then, we numerically solve four self-consistent/mean-field equations mentioned in the main text for each characteristic wavevector $\boldsymbol{Q}=(Q_{x},Q_{y})$. After finding solutions ($V_{0},V_{1},\mu,\lambda$) for all $\boldsymbol{Q}$ in the Brillouin zone, we compare their free energy density $F$ and the one with lowest $F$ is considered as the true solution. For example, we show free energy density $F$ versus $\boldsymbol{Q}$ for half-filling case with $J=4$ in Fig.~\ref{fig:6}. It is seen that $\boldsymbol{Q}=(\pi,\pi)$ has the lowest free energy while $\boldsymbol{Q}=(0,\pi)$ has rather high free energy density. Additionally, the particle-hole symmetry of half-filled Kondo lattice model on square lattice with nearest-neighbor-hopping requires $\mu=\lambda=0$. This fact can be proved as follows. One consider the real space version of Kondo lattice model (before mean-field decoupling and the constant part $-\sum_{i}\lambda_{i}$ is neglected),
\begin{widetext}
\begin{eqnarray}
H=-t\sum_{\langle i,j\rangle,\sigma}c_{i\sigma}^{\dag}c_{j\sigma}-\mu\sum_{i,\sigma}c_{i\sigma}^{\dag}c_{i\sigma}
+\sum_{i,\sigma}\lambda_{i}f_{i\sigma}^{\dag}f_{i\sigma}+\frac{J}{2}\sum_{i}\left(c^{\dag}_{i\uparrow}c_{i\downarrow}f^{\dag}_{i\downarrow}f_{i\uparrow}+c^{\dag}_{i\downarrow}c_{i\uparrow}f^{\dag}_{i\uparrow}f_{i\downarrow}+\frac{1}{2}(c^{\dag}_{i\uparrow}c_{i\uparrow}-c^{\dag}_{i\downarrow}c_{i\downarrow})(f^{\dag}_{i\uparrow}f_{i\uparrow}-f^{\dag}_{i\downarrow}f_{i\downarrow})\right).\nonumber
\end{eqnarray}
\end{widetext}
Here, the Kondo interaction term is written explicitly. Then, use the particle-hole transformation (similar to the Hubbard model)
\begin{eqnarray}
&&c_{i\uparrow}\rightarrow c_{i\uparrow}^{\dag},~~~~c_{i\downarrow}\rightarrow (-1)^{R_{i}}c_{i\downarrow}^{\dag}\nonumber\\
&&f_{i\uparrow}\rightarrow f_{i\uparrow}^{\dag},~~~~f_{i\downarrow}\rightarrow (-1)^{R_{i}}f_{i\downarrow}^{\dag},\nonumber
\end{eqnarray}
we find the transformed Hamiltonian reads as
\begin{widetext}
\begin{eqnarray}
\widetilde{H}=-t\sum_{\langle i,j\rangle,\sigma}c_{i\sigma}^{\dag}c_{j\sigma}+\mu\sum_{i,\sigma}c_{i\sigma}^{\dag}c_{i\sigma}
-\sum_{i,\sigma}\lambda_{i}f_{i\sigma}^{\dag}f_{i\sigma}+\frac{J}{2}\sum_{i}\left(c^{\dag}_{i\uparrow}c_{i\downarrow}f^{\dag}_{i\downarrow}f_{i\uparrow}+c^{\dag}_{i\downarrow}c_{i\uparrow}f^{\dag}_{i\uparrow}f_{i\downarrow}+\frac{1}{2}(c^{\dag}_{i\uparrow}c_{i\uparrow}-c^{\dag}_{i\downarrow}c_{i\downarrow})(f^{\dag}_{i\uparrow}f_{i\uparrow}-f^{\dag}_{i\downarrow}f_{i\downarrow})\right),\nonumber
\end{eqnarray}
\end{widetext}
which means the Kondo interaction term is unchanged under the particle-hole transformation but the chemical potential and Lagrangian multiplier parts acquire minus sign ($\mu\rightarrow-\mu,\lambda_{i}\rightarrow-\lambda_{i}$). Therefore, the particle-hole symmetry is conserved ($H=\widetilde{H}$) only if $\mu=\lambda_{i}=0$.

We note that $\boldsymbol{Q}=(0.8\pi,\pi)$ is an example of $(q,\pi)$ regime in Fig.~\ref{fig:2}
while $(0,0.6\pi)$ is a generic point in the regime with $\boldsymbol{Q}=(0,q)$.

\begin{table}[h!]
\begin{center}
\caption{Typical data used in phase diagram Fig.~\ref{fig:2}}\label{table_1}
\begin{tabular}{|l|c|c|c|c|c|c|}
\hline
$Q=(Q_{x},Q_{y})$ & J & \textbf{$n_{c}$} & $V_{0}$& $V_{1}$& $\mu$ &  $\lambda$\\
\hline
$(\pi,\pi)$    & 4   & 1    & 0.4556 & 0.0121 & 0.0000  & 0.0000\\

$(0.8\pi,\pi)$ & 0.5 & 0.7  & 0.0127 & 0.0039 & -0.7213 & -0.0275 \\

$(0,\pi)$      & 2   & 0.25 & 0.1355 & 0.0490 & -2.6481 & -0.0773 \\

$(0,0.6\pi)$   & 1   & 0.2    & 0.0246 & 0.0145 & -2.8365 & -0.0358 \\
\hline
\end{tabular}
\end{center}
\end{table}
\subsection{Green function, spectral function and local density of state}
\begin{figure}
\includegraphics[width=0.98\linewidth]{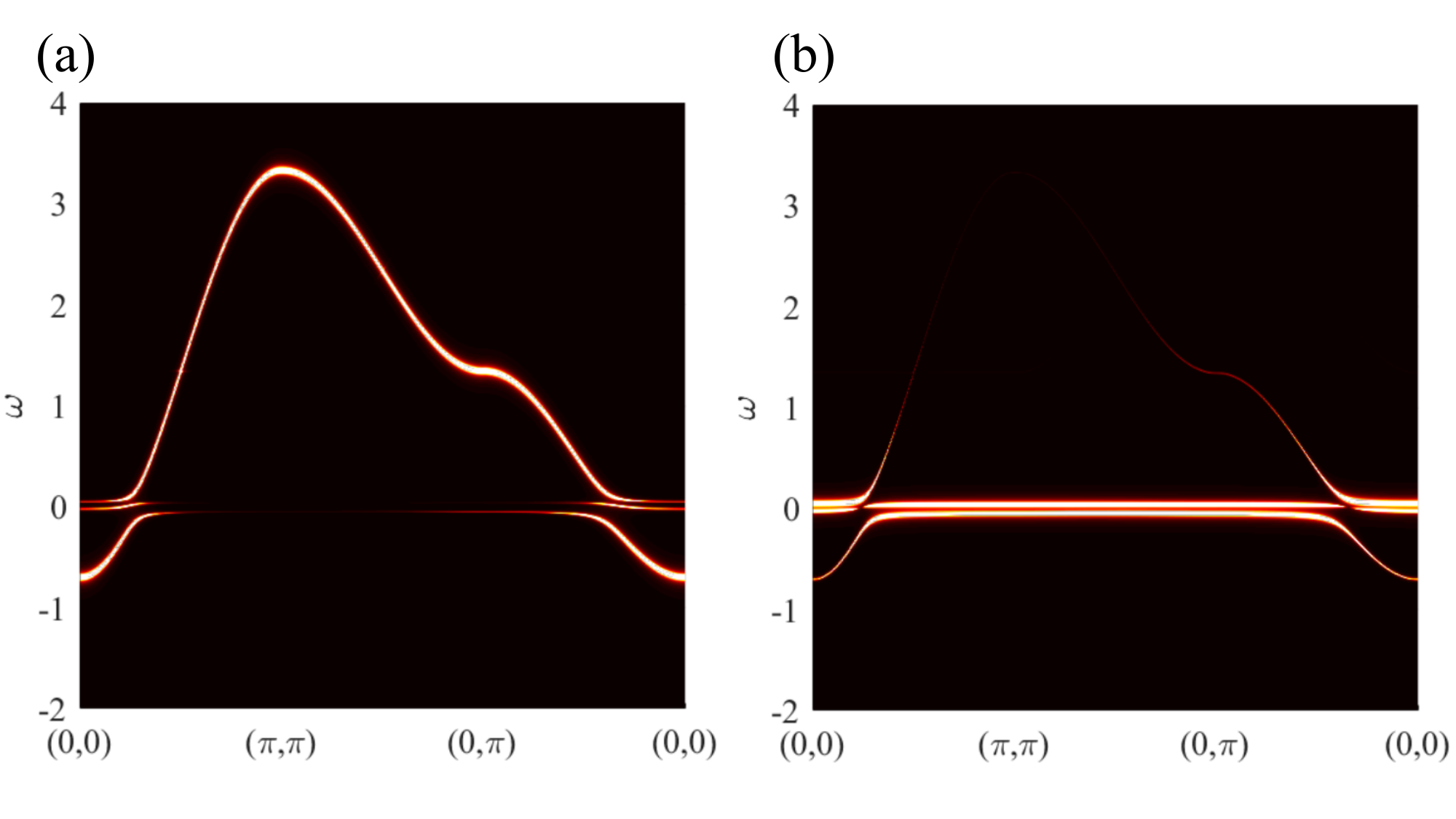}
\caption{\label{fig:7} The spectral function of conduction electron $A_{cc}(k,\omega)$ (a) and the Abrikosov fermion $A_{ff}(k,\omega)$ (b) for KHW phase with $\boldsymbol{Q}=(0,\pi)$.}
\end{figure}

The single-particle Green function can be found by $G=(\omega-H_{k})^{-1}$, so the single-particle Green function for conduction electron and Abrikosov fermion are $G_{11}(k,\omega)=G_{cc}(k,\omega),~~G_{44}(k,\omega)=G_{ff}(k,\omega)$. From $G_{cc}(k,\omega),G_{ff}(k,\omega)$, we can obtain their spectral function via $A_{cc}(k,\omega)=-\frac{1}{\pi}\mathrm{Im}G_{cc}(k,\omega),A_{ff}(k,\omega)=-\frac{1}{\pi}\mathrm{Im}G_{ff}(k,\omega)$. (see Fig.~\ref{fig:7} for an example)

We also examine the system's density of state, like $N_{cc}(\omega)=\frac{1}{N_{s}}\sum_{k}A_{cc}(k,\omega)$ and $N_{ff}(\omega)=\frac{1}{N_{s}}\sum_{k}A_{ff}(k,\omega)$. It is interesting to see the spatial dependence of density of state, the local density of state $N_{cc}(i,\omega)=-\frac{1}{\pi}\mathrm{Im}G_{cc}(i,\omega)$ and $N_{ff}(i,\omega)=-\frac{1}{\pi}\mathrm{Im}G_{ff}(i,\omega)$. We may calculate $G_{cc}(i,\omega)=\langle\langle c_{i\sigma}|c_{i\sigma}^{\dag}\rangle\rangle$ and $G_{ff}(i,\omega)=\langle\langle f_{i\sigma}|f_{i\sigma}^{\dag}\rangle\rangle$, where
\begin{equation}
\langle\langle c_{i\sigma}|c_{i\sigma}^{\dag}\rangle\rangle=-i\int_{-\infty}^{\infty}dte^{i(\omega+i0^{+})t}\theta(t)\langle[c_{i\sigma}(t),c_{i\sigma}^{\dag}]_{+}\rangle.\nonumber
\end{equation}
Use Fourier transformation, we find
\begin{widetext}
\begin{eqnarray}
G_{cc}(i,\omega)&=&\frac{1}{N_{s}}\sum_{k,k'}e^{ikR_{i}}e^{-ik'R_{i}}\langle\langle c_{k\sigma}|c_{k'\sigma}^{\dag}\rangle\rangle
=\frac{1}{N_{s}}\sum_{k,k'}e^{ikR_{i}}e^{-ik'R_{i}}(\delta_{k',k}\langle\langle c_{k\sigma}|c_{k\sigma}^{\dag}\rangle\rangle+\delta_{k',k+Q}\langle\langle c_{k\sigma}|c_{k+Q\sigma}^{\dag}\rangle\rangle)\nonumber\\
&=&\frac{1}{N_{s}}\sum_{k}\delta_{k',k}\langle\langle c_{k\sigma}|c_{k\sigma}^{\dag}\rangle\rangle
+\frac{1}{N_{s}}\sum_{k}e^{-i\boldsymbol{Q}\cdot\boldsymbol{R_{i}}}\langle\langle c_{k\sigma}|c_{k+Q\sigma}^{\dag}\rangle\rangle.\nonumber
\end{eqnarray}
\end{widetext}
Thus,
\begin{equation}
G_{cc}(i,\omega)=\frac{1}{N_{s}}\sum_{k}(G_{11}(k,\omega)+e^{-i\boldsymbol{Q}\cdot \boldsymbol{R_{i}}}G_{13}(k,\omega)).
\end{equation}
Similarly,
\begin{equation}
G_{ff}(i,\omega)=\frac{1}{N_{s}}\sum_{k}(G_{44}(k,\omega)+e^{-i\boldsymbol{Q}\cdot \boldsymbol{R_{i}}}G_{42}(k,\omega)).
\end{equation}
We also have
\begin{equation}
G_{cf}(i,\omega)=\frac{1}{N_{s}}\sum_{k}(G_{14}(k,\omega)+e^{-i\boldsymbol{Q}\cdot \boldsymbol{R_{i}}}G_{12}(k,\omega)),
\end{equation}
and
\begin{equation}
G_{fc}(i,\omega)=\frac{1}{N_{s}}\sum_{k}(G_{41}(k,\omega)+e^{-i\boldsymbol{Q}\cdot \boldsymbol{R_{i}}}G_{43}(k,\omega)).
\end{equation}
\subsection{The half-filling case with \texorpdfstring{$\boldsymbol{Q}=(\pi,\pi)$}{E=mc2} and others}

\begin{figure}
\includegraphics[width=0.95\linewidth]{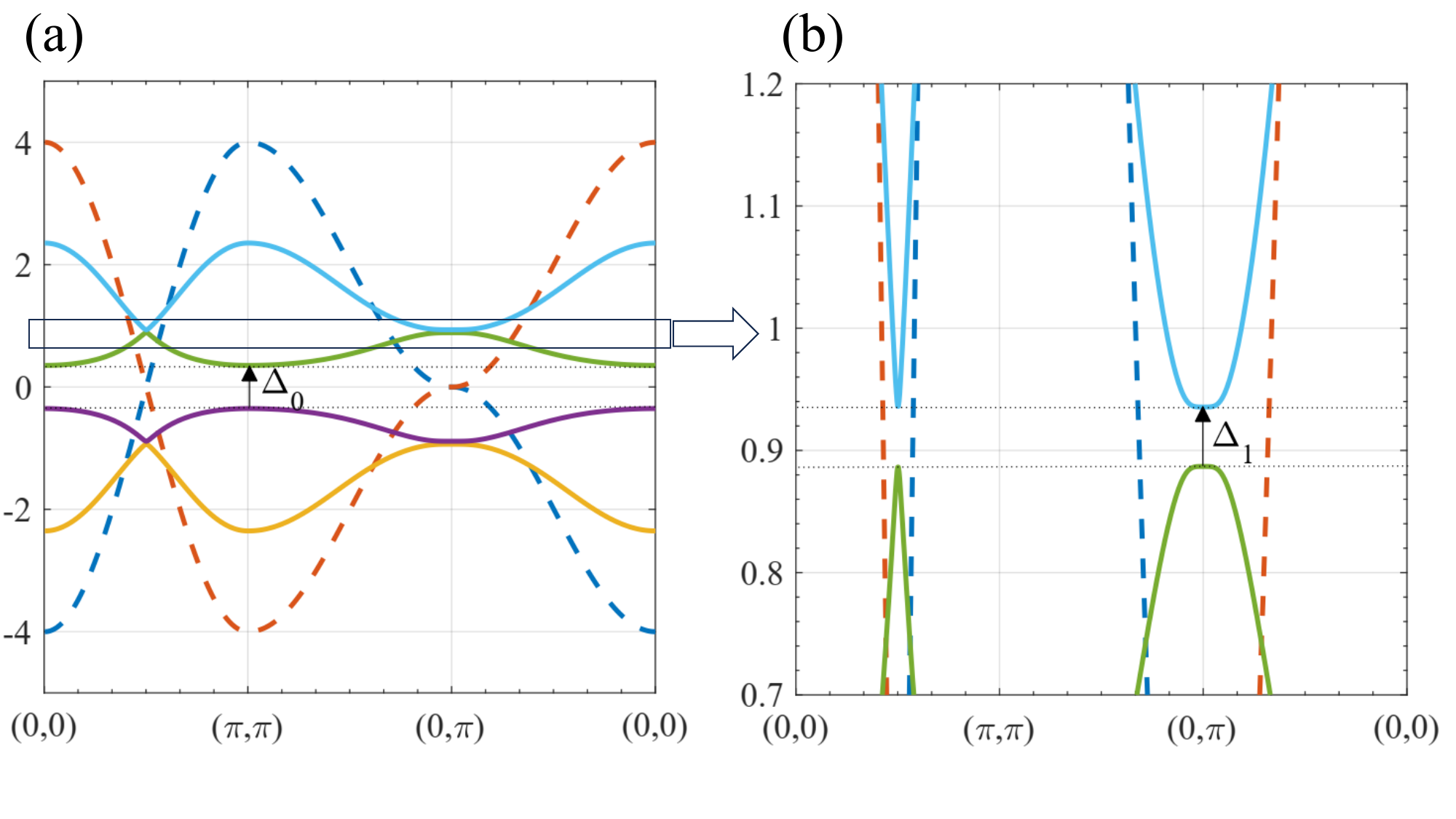}
\caption{\label{fig:8} The quasi-particle bands for half-filling case with $J=4$ with $Q=(\pi,\pi)$. The hybridization gap $\Delta_{0},\Delta_{1}$ exist around Fermi energy and high energy regime.}
\end{figure}
\begin{figure}
\includegraphics[width=0.95\linewidth]{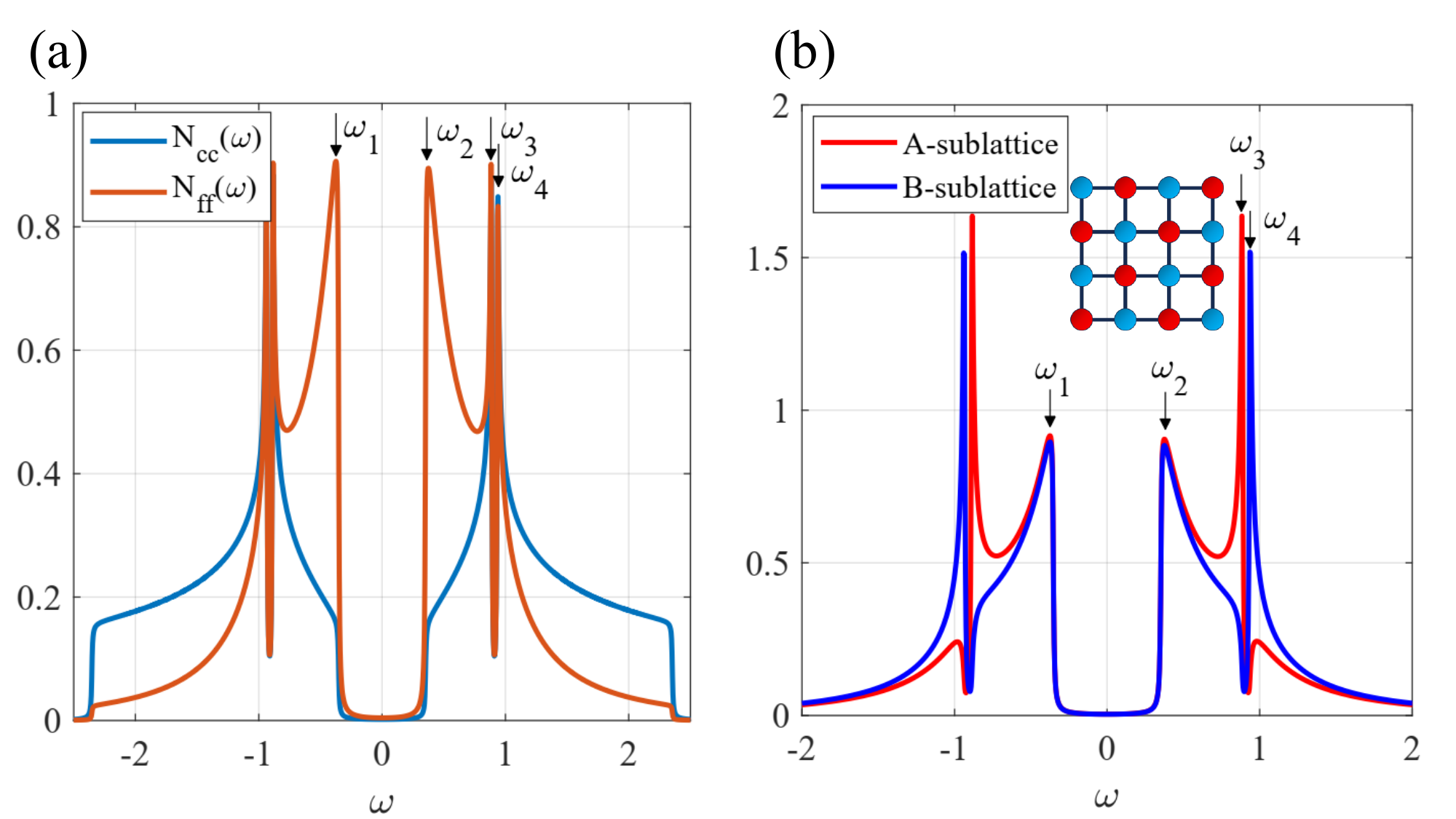}
\caption{\label{fig:9} (a) The DOS of conduction electron $N_{cc}(\omega)$ and Abrikosov fermion $N_{ff}(\omega)$ and (b) LDOS of Abrikosov fermion on A and B sublattice for KHW phase with $\boldsymbol{Q}=(\pi,\pi)$.}
\end{figure}

We consider the half-filling case with $J=4$ and $\boldsymbol{Q}=(\pi,\pi)$. From Fig.~\ref{fig:8}, there are four hybridized bands as seen in the case of $\boldsymbol{Q}=(0,\pi)$ discussed in the main text. One finds a large gap $\Delta_{0}\sim0.765$ around Fermi energy and a smaller gap $\Delta_{1}\sim0.049$ at high energy regime. Note that these two ones are the indirect gap, which is crucial for thermodynamics and spectrum. ($V_{0}$ and $V_{1}$ may be seen as the direct gap and response for transport properties like optic conductivity) If we use simple formula $\frac{(JV_{0})^{2}}{4t},\frac{(JV_{1})^{2}}{4t}$ to estimate $\Delta_{0},\Delta_{1}$, we find the estimated  $\Delta_{0}\sim0.832,\Delta_{1}\sim0.0006$. ($V_{0}=0.456,V_{1}=0.012$ from solution of self-consistent equations) Thus, the estimated values of small Kondo hybridization are much smaller than the ones extracted from hybridized bands, which means the estimation formula is seems to be crude. Furthermore, Fig.~\ref{fig:9}(a) shows DOS for both conduction electron and Abrikosov fermion, where the location of peaks in DOS is consistent with nearly flat-band regime in Fig.~\ref{fig:8}. It is seen that $\omega_{2}-\omega_{1}=\Delta_{0}$ and $\omega_{4}-\omega_{3}=\Delta_{1}$.

From peaks of LDOS in Fig.~\ref{fig:9}(b), we find both A and B sublattice have the same hybridization gap $\Delta_{0}$,
however, the small gap $\Delta_{1}$ is only visible for A sublattice site while we cannot extract it for B sublattice since it only has one peak in its LDOS.

Before ending this section, we note that the cases with generic incommensurate $Q$ (like ones in Table~\ref{table_1}) have similar band structure with $\boldsymbol{Q}=(0,\pi)$. For $\boldsymbol{Q}=(0,0.6\pi)$ and $(0.8\pi,\pi)$, because their $V_{0},V_{1}$ are too small, the (bare) conduction electron and localized Abrikosov fermion are nearly decoupled, thus the corresponding DOS and LDOS is trivial.
\subsection{The case for $\boldsymbol{Q}=(0,0)$}
\begin{figure}
\includegraphics[width=0.95\linewidth]{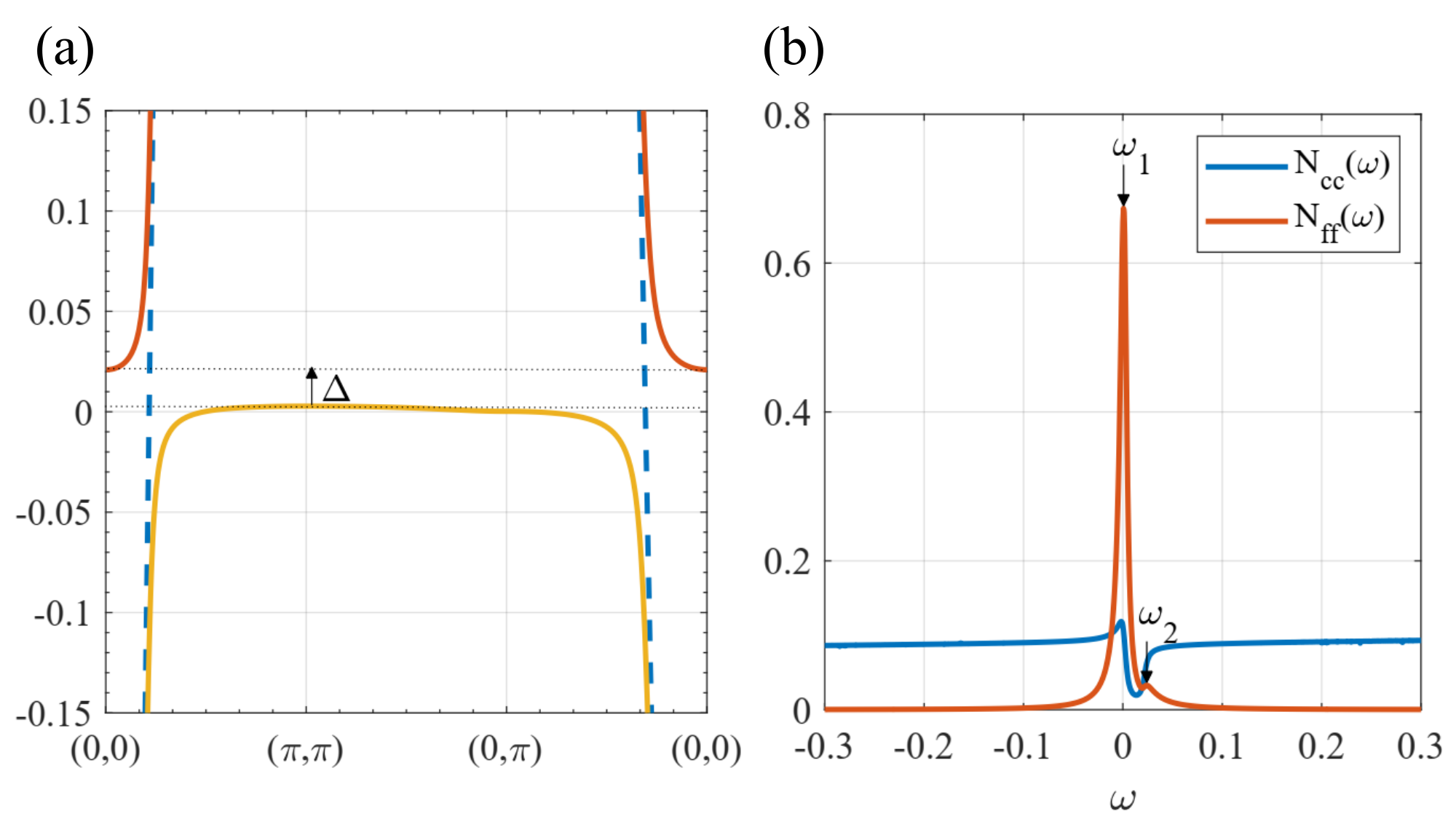}
\caption{\label{fig:10} (a) The two quasi-particle bands $E_{k\pm}$, (b) the DOS of conduction electron $N_{cc}(\omega)$ and Abrikosov fermion $N_{ff}(\omega)$ for the spatially uniform Kondo hybridization phase with $\boldsymbol{Q}=(0,0)$.}
\end{figure}

When the ordering wavevector $\boldsymbol{Q}=(0,0)$, the Hamiltonian in the main text is simplified into
\begin{eqnarray}
H=\sum_{k\sigma}\Psi_{k\sigma}^{\dag}H_{k}\Psi_{k\sigma}+N_{s}\left[2J(V_{0}^{2}+V_{1}^{2}+2V_{0}V_{1})-\lambda\right]\nonumber
\end{eqnarray}
with $\Psi_{k\sigma}^{\dag}=(c^{\dag}_{k\sigma}, f^{\dag}_{k\sigma}, c^{\dag}_{k\sigma}, f^{\dag}_{k\sigma})$ and
\begin{equation}
H_{k}=\frac{1}{2}\left(
                          \begin{array}{cccc}
                            \varepsilon_{k} & -JV_{1} & 0 & -JV_{0} \\
                            -JV_{1} & 0 & -JV_{0} & \lambda \\
                            0 & -JV_{0} & \varepsilon_{k} & -JV_{1} \\
                            -JV_{0} & \lambda & -JV_{1} & 0 \\
                          \end{array}
                        \right),\nonumber
\end{equation}
which leads to
\begin{widetext}
\begin{eqnarray}
H=\sum_{k\sigma}\left(
                  \begin{array}{cc}
                    c^{\dag}_{k\sigma} & f^{\dag}_{k\sigma} \\
                  \end{array}
                \right)
\left(
                  \begin{array}{cc}
                    \varepsilon_{k} & -J(V_{0}+V_{1}) \\
                    -J(V_{0}+V_{1}) & \lambda \\
                  \end{array}
                \right)\left(
                         \begin{array}{c}
                           c_{k\sigma} \\
                           f_{k\sigma} \\
                         \end{array}
                       \right)+N_{s}\left[2J(V_{0}+V_{1})^{2}-\lambda\right].
\end{eqnarray}
\end{widetext}
This Hamiltonian is the familiar one encountered in spatially uniform Kondo hybridization state. One finds two quasiparticle bands
$E_{k\pm}=\frac{1}{2}(\varepsilon_{k}+\lambda\pm\sqrt{(\varepsilon_{k}-\lambda)^{2}+4J^{2}(V_{0}+V_{1})^{2}})$. Therefore, only one effective hybridization $V=V_{0}+V_{1}$ is active for $Q=(0,0)$ case. The corresponding free energy density is given by
\begin{equation}
F=-2\frac{T}{N_{s}}\sum_{k}\left(\ln(1+e^{-\beta E_{k+}})+\ln(1+e^{-\beta E_{k-}})\right)+2JV^{2}-\lambda,
\end{equation}
therefore the self-consistent equations are
\begin{widetext}
\begin{eqnarray}
&&\frac{\partial F}{\partial \lambda}=0\Rightarrow 1=\frac{2}{N_{s}}\sum_{k}\left(f_{F}(E_{k+})\frac{\partial E_{k+}}{\partial\lambda}+f_{F}(E_{k-})\frac{\partial E_{k-}}{\partial\lambda}\right),\nonumber\\
&&\frac{\partial F}{\partial V^{2}}=0\Rightarrow -2J=\frac{2}{N_{s}}\sum_{k}\left(f_{F}(E_{k+})\frac{\partial E_{k+}}{\partial V^{2} }+f_{F}(E_{k-})\frac{\partial E_{k-}}{\partial V^{2}}\right),\nonumber\\
&&n_{c}=-\frac{\partial F}{\partial \mu}=-\frac{2}{N_{s}}\sum_{k}\left(f_{F}(E_{k+})\frac{\partial E_{k+}}{\partial \mu }+f_{F}(E_{k-})\frac{\partial E_{k-}}{\partial \mu}\right),
\end{eqnarray}
\end{widetext}
where
\begin{equation}
\frac{\partial E_{k\pm}}{\partial\lambda}=\frac{1}{2}\left(1\mp\frac{\varepsilon_{k}-\lambda}{\sqrt{(\varepsilon_{k}-\lambda)^{2}+4J^{2}V^{2}}}\right),\nonumber
\end{equation}
\begin{equation}
\frac{\partial E_{k\pm}}{\partial V^{2}}=\pm\frac{J^{2}}{\sqrt{(\varepsilon_{k}-\lambda)^{2}+4J^{2}V^{2}}},\nonumber
\end{equation}
\begin{equation}
\frac{\partial E_{k\pm}}{\partial \mu}=\frac{1}{2}\left(-1\mp\frac{\varepsilon_{k}-\lambda}{\sqrt{(\varepsilon_{k}-\lambda)^{2}+4J^{2}V^{2}}}\right).\nonumber
\end{equation}
The single-particle Green functions for conduction electron and Abrikosov fermion are
\begin{equation}
G_{cc}(k,\omega)=\frac{1}{\omega-E_{k+}}\frac{E_{k+}-\lambda}{E_{k+}-E_{k-}}-\frac{1}{\omega-E_{k-}}\frac{E_{k-}-\lambda}{E_{k+}-E_{k-}}\nonumber
\end{equation}
and
\begin{equation}
G_{ff}(k,\omega)=\frac{1}{\omega-E_{k+}}\frac{E_{k+}-\varepsilon_{k}}{E_{k+}-E_{k-}}-\frac{1}{\omega-E_{k-}}\frac{E_{k-}-\varepsilon_{k}}{E_{k+}-E_{k-}}.\nonumber
\end{equation}
So, the DOS should be
\begin{equation}
N_{cc}(\omega)=\frac{1}{N_{s}}\sum_{k}\left(-\frac{1}{\pi}\mathrm{Im}G_{cc}(k,\omega)\right),
\end{equation}
\begin{equation}
N_{ff}(\omega)=\frac{1}{N_{s}}\sum_{k}\left(-\frac{1}{\pi}\mathrm{Im}G_{ff}(k,\omega)\right).
\end{equation}
For $J=2$ and $n_{c}=0.15$, we find $V=V_{0}+V_{1}=0.0599$ and $\mu=-3.1335,\lambda=0.0048$. Fig.~\ref{fig:10}(a) illustrates the quasi-particle bands $E_{k\pm}$ of the spatially uniform Kondo hybridization phase with $\boldsymbol{Q}=(0,0)$ and only one hybridization gap $\Delta\sim0.017$ exists. Fig.~\ref{fig:10}(b) shows the corresponding DOS and we see $\omega_{2}-\omega_{1}=\Delta$. Amusingly, we note the DOS near Fermi energy for conduction electron looks like the Fano line-shape, which is usually the result of interference between conduction electron and local moment of $f$-electron\cite{Morr2017}.
\subsection{The Fano line-shape of KHW}
The site-resolved STM spectrum, including the tunnelling into conduction electron and local moment has the following form\cite{Morr2010},
\begin{widetext}
\begin{equation}
N_{STM}(i,\omega)=t_{c}^{2}N_{cc}(i,\omega)+t_{f}^{2}N_{ff}(i,\omega)+t_{c}t_{f}\left(N_{cf}(i,\omega)+N_{fc}(i,\omega)\right),
\end{equation}
\end{widetext}
where $t_{c}, t_{f}$ are amplitudes of tip electron tunneling into conduction electron and local moment orbit. $N_{cc}(i,\omega),N_{ff}(i,\omega),N_{cf}(i,\omega),N_{fc}(i,\omega)$ are LDOS. One note that besides the independent contribution of conduction electron and Abrikosov fermion, there are terms $N_{cf}(i,\omega),N_{fc}(i,\omega)$ involving their hybridization.

One usually use the following fitting formula to analyze the STM spectrum\cite{Yu2026,Wolfle2010},
\begin{equation}
N_{STM}(\omega)\propto d\frac{(q+(\omega-\epsilon_{0})/\Gamma)^2}{1+((\omega-\epsilon_{0})/\Gamma)^2}+c,~~~~
\end{equation}
where $\epsilon_{0}$ is the resonance energy, $\Gamma$ is the width of Kondo resonance and $q$ is the $f-c$ tunneling probability ratio (also called Fano factor). $c,d$ are trivial factors. We note that above formula is originally derived from single-impurity Kondo problem. For Kondo lattice systems studied here, one expect $\Gamma\sim\Delta$ and $\epsilon_{0}\sim\lambda$. Thus, we can extract $\epsilon_{0},\Gamma,q$ or $\lambda,\Delta,q$ by fitting to experimental data. If the system itself is not uniform as the cases for KHW state, these parameters are also site-resolved, i.e. $\epsilon_{0}(i),\Gamma(i),q(i)$.

For the KHW state with $\boldsymbol{Q}=(0,\pi),J=2,n_{c}=0.25$, it has two-sublattice structure ($A$ and $B$ sublattice) and the fitting parameters for each sublattice are shown in Table.~\ref{table_2}. (We choose $t_{f}/t_{c}=0.1$.) A large damping factor $\delta=0.05$ is used, otherwise, the spectrum will strongly deviate from Fano line-shape and the fitting formula above may be meaningless.

\begin{table}[h!]
\begin{center}
\caption{Fitting parameters extracted for STM spectrum of KHW state with $\boldsymbol{Q}=(0,\pi),J=2,n_{c}=0.25$.}\label{table_2}
\begin{tabular}{|l|c|c|c|c|c|r|}
\hline
site-type & \textbf{$q$} & \textbf{$\epsilon_{0}$} & $\Gamma$ & $c$ & $d$\\
\hline
A sublattice    & 2.9290   & -0.0098    & 0.0436 & 0.1691 & 0.0196  \\

B sublattice & 0.3808 & -0.0677  & 0.0533 & 0.3861 & -0.2003\\
\hline
\end{tabular}
\end{center}
\end{table}

\begin{thebibliography}{58}%
\bibitem{Hewson1993} A. Hewson, \emph{The Kondo Problem to Heavy Fermions} (Cambridge
University Press, Cambridge, 1993).
\bibitem{Doniach1977}
S. Doniach, The Kondo lattice and weak antiferromagnetism, Physica B+C \textbf{91}, 231 (1977).

\bibitem{Lohneysen2007} H. v. L\"{o}hneysen, A. Rosch, M. Vojta, and P. W\"{o}lfle, Fermiliquid instabilities at magnetic quantum phase transitions, Rev. Mod. Phys. \textbf{79}, 1015 (2007).

\bibitem{Coleman2015} P. Coleman, \emph{Introduction to Many Body Physics} (Cambridge
University Press, Cambridge, 2015).
\bibitem{Coleman2001}
P. Coleman, C. P\'{e}pin, Q. Si and R. Ramazashvili, How do Fermi liquids get heavy and die?, J. Phys. Condens. Matter \textbf{13}, R723 (2001).
\bibitem{Si2001}
Q. Si, S. Rabello, K. Ingersent and J. L. Smith, Nature (London) \textbf{413}, 804 (2001).
\bibitem{Senthil2004}
T. Senthil, M. Vojta and S. Sachdev, Weak magnetism and non-Fermi liquids near heavy-fermion critical points, Phys. Rev. B \textbf{69}, 035111 (2004).
\bibitem{Vojta2010}
M. Vojta, Orbital-Selective Mott Transitions: Heavy Fermions and Beyond, J Low Temp Phys \textbf{161}, 203 (2010).


\bibitem{Yu2026} X. Yu, S. Yu, Z. Wu, A. G. Eaton, A. Cabala, M.
Vali\v{s}ka, J. Li, R. Zhou, Y.-f. Yang, Z. Wang, P. Sun and R. Wu, Observation of Kondo hybridization wave in UTe$_{2}$, arXiv:2603:10552.
\bibitem{Cao2026} L. Cao, J. Shi, L. Liu, X. Luo, Y.-P. Sun, Y.-f. Yang, Y. Yao, J. Mao and Y. Jiang, Discovery of a hybridization-wave electronic order in a van der Waals Kondo lattice, arXiv:2603.12720.
\bibitem{Dubi2011}
Y. Dubi and A. V. Balatsky, Hybridization Wave as the ''Hidden Order'' in URu$_{2}$Si$_{2}$, Phys. Rev. Lett. \textbf{106}, 086401 (2011)
\bibitem{Su2011}
J.-J. Su, Y. Dubi, P. Wolfle and A. V. Balatsky, A charge density wave in the hidden order
state of URu$_{2}$Si$_{2}$, J. Phys.: Condens. Matter \textbf{23}, 094214 (2011).
\bibitem{Xie2017}
N. Xie, D. Hu and Y.-f. Yang, Hybridization oscillation in the onedimensional Kondo-Heisenberg
model with Kondo holes, Sci. Rep. \textbf{7}, 11924 (2017).
\bibitem{Schmidt2010}
A. R. Schmidt, M. H. Hamidian, P. Wahl, F. Meier, A. V. Balatsky, J. D. Garrett, T. J. Williams, G. M. Luke and J. C. Davis, Imaging the Fano lattice to 'hidden order' transition in URu$_{2}$Si$_{2}$, Nature(London) \textbf{465}, 570 (2010).

\bibitem{Mydosh2011}
J. A. Mydosh and P. M. Oppeneer, Colloquium: Hidden order, superconductivity, and magnetism:
The unsolved case of URu$_{2}$Si$_{2}$, Rev. Mod. Phys. \textbf{83}, 1301 (2011)

\bibitem{Morr2017}
D. K. Morr, Theory of scanning tunneling spectroscopy: from Kondo impurities to heavy fermion materials, Rep. Prog. Phys. \textbf{80}, 014502 (2017).
\bibitem{Yang2009}
Y.-f. Yang, Fano effect in the point contact spectroscopy of heavy-electron materials, Phys. Rev. B \textbf{79}, 241107(R) (2009)
\bibitem{Maltseva2009}
M. Maltseva, M. Dzero and P. Coleman, Electron Cotunneling into a Kondo Lattice, Phys. Rev. Lett. \textbf{103}, 206402 (2009).
\bibitem{Morr2010}
J. Figgins and D. K. Morr, Differential Conductance and Quantum Interference in Kondo Systems, Phys. Rev. Lett. \textbf{104}, 187202 (2010).
\bibitem{Wolfle2010}
P. W\"{o}lfle, Y. Dubi and A. V. Balatsky, Tunneling into Clean Heavy Fermion Compounds: Origin of the Fano Line Shape, Phys. Rev. Lett. \textbf{105}, 246401 (2010).

\bibitem{Tsunetsugu1997} H. Tsunetsugu, M. Sigrist and K. Ueda, The ground-state phase diagram of the one-dimensional Kondo lattice model, Rev. Mod. Phys. \textbf{69}, 809 (1997).

\bibitem{Ghaemi2007} P. Ghaemi and T. Senthil, Higher angular momentum Kondo liquids, Phys. Rev. B \textbf{75}, 144412 (2007).
\bibitem{Weber2008} H. Weber and M. Vojta, Heavy-fermion metals with hybridization nodes:
Unconventional Fermi liquids and competing phases, Phys. Rev. B \textbf{77}, 125118 (2008).
\bibitem{Coleman1998} P. Coleman and A. M. Tsvelik, Local moments in an interacting environment, Phys. Rev. B \textbf{57}, 12757 (1998).
\bibitem{Lacroix1979} C. Lacroix and M. Cyrot, Phase diagram of Kondo lattice, Phys. Rev. B \textbf{20}, 1969 (1979).
\bibitem{Zhang2011} G.-M. Zhang, Y.-H. Su and L. Yu, Lifshitz transitions in a heavy Fermi liquid driven by short-range antiferromagnetic correlations in the two-dimensional Kondo lattice model, Phys. Rev. B \textbf{83}, 033102 (2011).
\bibitem{Irkhin2023}
V. Y. Irkhin, Magnetism, Non-Fermi-Liquid Behavior, and Deconfinement in Kondo Lattices, JETP Letters \textbf{118}, 45 (2023).
\bibitem{Costa2017}
N. de C. Costa, J. P. de Lima, R. R. dos Santos, Spiral magnetic phases on the Kondo Lattice Model: A Hartree-Fock approach, J. Magn. Magn. Mater. \textbf{423}, 74 (2017).
\bibitem{Chen2021}
C. Chen, I. Sodemann and P. A. Lee, Competition of spinon Fermi surface and heavy Fermi liquid states from the periodic
Anderson to the Hubbard model, Phys. Rev. B \textbf{103}, 085128 (2021).
\bibitem{Mahan2000} G. D. Mahan, \emph{Many-Particle Physics} (Kluwer Academic/Plenum Publishers, New York, 2000).
\bibitem{Nikolaenko2026}
A. Nikolaenko, R. Rende, L. L. Viteritti, S. Sachdev and Y.-H. Zhang, Fermi surface change and d-wave superconductivity
in the square lattice Kondo-Heisenberg model, arXiv:2606.23799.
\bibitem{Chen2022}
J. Chen, J. Wang, D. Hu and Y.-f. Yang, Continuous ferromagnetic quantum phase transition on an anisotropic Kondo lattice, Phys. Rev. B \textbf{106}, 075114 (2022).

\bibitem{Cohen2026}
E. Y. Cohen, F. F. Assaad and S. Gazit, Antiferromagnetism and Stripe Channel Order in the SU(N)-Symmetric Two-Channel Kondo Lattice Model, Phys. Rev. Lett. \textbf{137}, 036504 (2026).

\bibitem{Assaad1999} F. F. Assaad, Quantum Monte Carlo Simulations of the Half-Filled Two-Dimensional Kondo Lattice Model, Phys. Rev. Lett. \textbf{83}, 796 (1999).
\end{thebibliography}
\end{document}